\documentclass{tcibook}
\usepackage{fancyhea}
\usepackage{sidecap}

\usepackage{work}
\usepackage{bm}       
\usepackage{graphicx}
\usepackage{hyperref}      
\usepackage{lineno}

\usepackage{epsfig}
\usepackage{graphicx}
\usepackage{type1cm}
\usepackage{eso-pic}
\usepackage{color}

\usepackage{enumitem}
\usepackage[us,12hr]{datetime}

\setitemize[0]{leftmargin=*}

\newcommand{\nc}{\newcommand}  

\def\ie{{\it i.e.}}
\def\eg{{\it e.g.}}

\def\beq{\begin{equation}}
\def\eeq#1{\label{#1}\end{equation}}
\def\eeqn{\end{equation}}

\newenvironment{Eqnarray}%
   {\arraycolsep 0.14em\begin{eqnarray}}{\end{eqnarray}}
\def\beqa{\begin{Eqnarray}}
\def\eeqa#1{\label{#1}\end{Eqnarray}}
\def\eeqan{\end{Eqnarray}}

\nc{\ra}{\rightarrow}  
\nc{\slsh}{\slash\hspace*{-0.22cm}}
\def\Re{{\cal R \mskip-4mu \lower.1ex \hbox{\it e}\,}}
\def\Im{{\cal I \mskip-5mu \lower.1ex \hbox{\it m}\,}}

\nc{\vev}[1]{ \left\langle {#1} \right\rangle }
\nc{\bra}[1]{ \langle {#1} | }
\nc{\ket}[1]{ | {#1} \rangle }
\nc{\fb}{\,{\rm fb}^{-1}}
\nc{\ev}{{\rm eV}}
\nc{\kev}{{\rm keV}}
\nc{\Mev}{{\rm MeV}}
\nc{\gev}{{\rm GeV}}
\nc{\tev}{{\rm TeV}}
\nc{\mev}{{\rm MeV}}

\def\del{\partial}
\def\Dslash{\not{\hbox{\kern-4pt $D$}}}
\def\dslash{\not{\hbox{\kern-2pt $\del$}}}
\def\pslash{\not{\hbox{\kern-2pt $p$}}}
\def\ETmiss{ \not{\hbox{\kern-4pt $E$}}_T }

\def\ee{e^+e^-}

\def\msb{{\bar{\ssstyle M \kern -1pt S}}}

\newcommand{\eqref}[1]{Eq.~(\ref{#1})}

\newcommand{\Secref}[1]{Section~\ref{sec:#1}}
\newcommand{\Secsref}[2]{Sections~\ref{sec:#1} and \ref{sec:#2}}

\newcommand{\Figref}[1]{Figure~\ref{fig:#1}}

\newcommand{\alt}{\mathrel{\hbox{\rlap{\hbox{\lower4pt\hbox{$\sim$}}}\hbox{$<$}}}}
\newcommand{\agt}{\mathrel{\hbox{\rlap{\hbox{\lower4pt\hbox{$\sim$}}}\hbox{$>$}}}}

\def\beqn{\begin{eqnarray}} 
\def\eeqn{\end{eqnarray}} 
\def\be{\begin{equation}}
\def\ee{\end{equation}}

\begin{document}


\def\bibname{References}

\bibliographystyle{utphys}  

\raggedbottom

\pagenumbering{roman}

\parindent=0pt
\parskip=8pt
\setlength{\evensidemargin}{0pt}
\setlength{\oddsidemargin}{0pt}
\setlength{\marginparsep}{0.0in}
\setlength{\marginparwidth}{0.0in}
\marginparpush=0pt


\pagenumbering{arabic}

\renewcommand{\arraystretch}{1.25}
\addtolength{\arraycolsep}{-3pt}

\thispagestyle{empty}
\begin{centering}
\vfill

{\Huge\bf Planning the Future of U.S. Particle Physics}

{\Large \bf Report of the 2013 Community Summer Study}

\vfill

{\Huge \bf Chapter 4: Cosmic Frontier}

\vspace*{2.0cm}
{\Large \bf Conveners: J. L. Feng and S. Ritz}
\pagenumbering{roman}

\vfill

{\large  Study Conveners: M. Bardeen, W. Barletta, L.~A.~T.~Bauerdick, R. Brock,
D.~Cronin-Hennessy, M.~Demarteau, M.~Dine, J.~L. Feng, M. Gilchriese,
S. Gottlieb, J.~L.~Hewett, R. Lipton, H.~Nicholson, M.~E. Peskin,
S. Ritz, I.~Shipsey, H. Weerts}\\
\vspace{1cm}

{\large Division of Particles and Fields Officers in 2013:
J.~L. Rosner (chair), 
I. Shipsey (chair-elect), 
N. Hadley (vice-chair),
P. Ramond (past chair)}\\
\vspace{1cm}

{\large Editorial Committee:
R.~H. Bernstein,
N. Graf,
P. McBride,
M.~E. Peskin,
J.~L. Rosner,
N.~Varelas,
K. Yurkewicz}

\vfill

\end{centering}

\newpage
\mbox{\null}

\vspace{3.0cm}

{\Large \bf Authors of Chapter 4:}

\vspace{2.0cm}
 {\bf  J.~L.~Feng, S.~Ritz}, 
J.~J.~Beatty, 
J.~Buckley, 
D.~F.~Cowen, 
P.~Cushman, 
S.~Dodelson, 
C.~Galbiati, 
K.~Honscheid, 
D.~Hooper, 
M.~Kaplinghat, 
A.~Kusenko, 
K.~Matchev, 
D.~McKinsey, 
A.~E.~Nelson, 
A.~Olinto, 
S.~Profumo, 
H.~Robertson, 
L.~Rosenberg, 
G.~Sinnis,
T.~M.~P.~Tait

 \tableofcontents

\newpage

\pagenumbering{arabic}

\newcommand{\sfig}[2]{
\centering
\includegraphics[width=#2]{#1}
}
\newcommand{\Sfig}[2]{
    \begin{figure}[tb]
    \sfig{#1.pdf}{0.73\columnwidth}
\vspace*{.2in}
   \caption{#2} 
    \label{fig:#1}
    \end{figure}
}

\newcommand{\Sfigp}[2]{
    \begin{figure}[tbp]
    \sfig{#1.pdf}{0.83\columnwidth}
\vspace*{.1in}
   \caption{#2} 
    \label{fig:#1}
    \end{figure}
}

\setcounter{chapter}{3} 

\chapter{Cosmic Frontier}

\begin{center}\begin{boldmath}



\begin{center}

\medskip

\begin{large} {\bf Conveners: J.~L.~Feng and S.~Ritz} \end{large}


J.~J.~Beatty, 
J.~Buckley, 
D.~F.~Cowen, 
P.~Cushman, 
S.~Dodelson, 
C.~Galbiati, 
K.~Honscheid, 
D.~Hooper, 
M.~Kaplinghat, 
A.~Kusenko, 
K.~Matchev, 
D.~McKinsey, 
A.~E.~Nelson, 
A.~Olinto, 
S.~Profumo, 
H.~Robertson, 
L.~Rosenberg, 
G.~Sinnis,
T.~M.~P.~Tait
\end{center}




\end{boldmath}\end{center}




\section*{Executive summary}

\vspace*{-.1in}

Investigations at the Cosmic Frontier use the Universe as a laboratory
to learn about particle physics.  Our understanding of the Universe
has been transformed in recent years.  In particular, experiments at
the Cosmic Frontier have demonstrated that only 5\% of the contents of
the Universe are well understood, with the rest composed of mysterious
dark matter and dark energy.  As a result, the Cosmic Frontier now
plays a central role in the global particle physics program, providing
overwhelming evidence for new particles and new interactions, as well
as powerful, unique opportunities to address many of our most
fascinating questions: What is dark matter?  What is dark energy?  Why
is there more matter than antimatter?  What are the properties of
neutrinos?  How did the Universe begin?  What is the physics of the
Universe at the highest energies?

To identify outstanding scientific opportunities for the coming 10 to
20 years, the Cosmic Frontier Working Group was organized into six
subgroups: 1.~WIMP Dark Matter Direct Detection, 2.~WIMP Dark Matter
Indirect Detection, 3.~Non-WIMP Dark Matter, 4.~Dark Matter
Complementarity, 5.~Dark Energy and CMB, and 6.~Cosmic Particles and
Fundamental Physics.  In several cases, these subgroups were further
divided into topical working groups.  The work of these groups was
carried out through teleconferences and meetings, including the Cosmic
Frontier Workshop at SLAC, March 6--8, 2013
(http://www-conf.slac.stanford.edu/cosmic-frontier/2013), and the
Community Summer Study 2013 meeting in Minnesota, July 29--August 6,
2013 (http://www.hep.umn.edu/css2013).  More complete summaries and
references than can be presented here may be found in the talks from
these meetings, and the subgroup and topical group
summaries~\cite{Cushman:2013zza,Buckley:2013bha,Kusenko:2013saa,%
  Arrenberg:2013qaa,Dodelson:2013pln,Beatty:2013lza,%
  Cooley:2013mna,Weinberg:2013raj,Abazajian:2013vfg,Kim:2013ijr,%
  Abazajian:2013oma,Newman:2013cac,Huterer:2013xky,Rhodes:2013fyq,%
  Jain:2013wgs}.

The $\Lambda$CDM standard model of cosmology provides the backdrop for
much of Cosmic Frontier research.  In this model, the Universe
underwent a very early epoch of accelerated expansion (inflation),
which was followed by eras in which the Universe was dominated
successively by radiation, cold dark matter (CDM), and dark energy
($\Lambda$).  At present, the known particles make up only 5\% of the
energy density of the Universe, with neutrinos contributing at least
$0.1\%$.  The rest is 25\% dark matter and 70\% dark energy.
Remarkably, incisive measurements that explore all of the key
components of the model are now within reach.  The leaps in
sensitivity of the new facilities bring us to a time with strong
discovery potential in many areas.  Further surprises are likely in
this rapidly advancing area, with potentially far-reaching
consequences.

\vspace*{0.05in} {\bf Dark matter}

The work of Snowmass highlighted the coming decade as one of
particular promise for the goal of identifying dark matter.  Evidence
for particle dark matter has been building for 80 years through the
study of galaxy clusters, galactic rotation curves, weak lensing,
strong lensing, hot gas in galaxy clusters, galaxy cluster collisions,
supernovae, and the cosmic microwave background (CMB).  However, all
evidence so far is based on dark matter's gravitational interactions,
and its particle identity remains a deep mystery.

Among the many dark matter candidates, one well-known possibility is
weakly-interacting massive particles (WIMPs) with masses in the 1 GeV
to 100 TeV range.  Particles with these masses are strongly motivated
by particle physics, where they appear in many models designed to
address the gauge hierarchy problem (the great discrepancy between the
weak and Planck mass scales), and by cosmology, where they may obtain
the correct relic density either through thermal freeze-out or through
an asymmetry connecting their number density to that of baryons.

WIMP direct detection experiments search for the interactions of WIMPs
with normal matter.  WIMPs may scatter elastically off nuclei,
producing recoil energies in the 1--100 keV range, which can be
detected through phonons, ionization, scintillation, or other methods.
There are daunting backgrounds, and so direct detection experiments
must be placed deep underground.  In the last several years, however,
this field has seen a burgeoning of innovative approaches to
discriminate signal from background, including experiments
incorporating dual-phase media, self shielding, pulse shape
discrimination, and threshold detectors.

The first two decades of direct detection experiments have yielded a
diverse and successful program, resulting in ``Moore's Law''-type
progress, with sensitivities doubling roughly every 18 months.  In the
coming decade, this rate of progress is expected to continue or even
accelerate for both spin-independent and spin-dependent interactions.
Upcoming second generation (G2) experiments will improve sensitivities
by an order of magnitude, probing the Higgs-mediated cross sections
expected for well-known supersymmetric and extra-dimensional
candidates, and also extending the sensitivity to both $\sim \gev$
low-mass WIMPs, where possible signals have been reported, and $\sim
\tev$ masses that are beyond the reach of colliders. Following these
experiments, multi-ton-scale third generation (G3) experiments are
expected to improve current sensitivities by up to three orders of
magnitude and will either find dark matter or detect background events
from solar, atmospheric, and diffuse supernovae neutrinos.  Probing
beyond this sensitivity will require either background subtraction or
techniques such as directional detection or annual modulation.  The
Snowmass process produced a detailed census of present and proposed
direct detection facilities, with uniform treatment of their
capabilities and issues, along with a survey of promising
technologies.

WIMPs may also be found through indirect detection, in which pairs of
WIMPs annihilate, producing Standard Model particles, including gamma
rays, neutrinos, electrons and positrons, protons and antiprotons, and
deuterons and antideuterons.  Detection of these particles may be used
to constrain or infer dark matter properties.  The expectation that
WIMP annihilation in the early Universe determines the dark matter
abundance sets a natural velocity-averaged annihilation cross section
of $\langle \sigma_{\rm{an}} v \rangle \sim 3 \times 10^{-26}~{\rm
  cm}^3~{\rm s}^{-1}$ for indirect detection experiments.

Gamma rays from dark matter annihilation may be detected by both
space- and ground-based experiments.  In space, the Fermi-LAT has
recently demonstrated the promise of this approach, excluding the
natural cross section $\langle \sigma_{\rm{an}} v \rangle$ for dark
matter masses below 30 GeV, given certain halo profile and
annihilation channel assumptions, and the reach is expected to be
extended significantly with additional data.  On the ground, VERITAS
and other atmospheric Cherenkov telescopes have set significant limits
by looking for gamma rays from dark matter-rich dwarf galaxies.
Moving forward, the atmospheric Cherenkov telescope community has
coalesced to build the Cherenkov Telescope Array (CTA), with
sensitivity at the natural cross-section scale for dark matter masses
from 100 GeV to 10 TeV, far beyond current or planned colliders, for
conservative halo profiles and many of the possible annihilation
channels.  These projections require U.S. involvement in CTA, which
will double the planned mid-sized telescope array and enable critical
improvements in sensitivity and angular resolution.

Neutrinos also provide promising means for indirect detection of dark
matter.  High-energy neutrinos from the core of the Sun would be a
smoking-gun signal of dark matter particle annihilation.  The signal
depends primarily on the spin-dependent WIMP-nucleon scattering cross
section, which determines the capture rate.  Current bounds from
SuperK and IceCube already provide leading limits on this cross
section, and the Precision IceCube Next Generation Upgrade (PINGU), an
infill array upgrade to IceCube, will extend the sensitivity to lower
masses.  In the coming decade, IceCube and PINGU, along with
Hyper-Kamiokande, will probe cross sections one to two orders of
magnitude below current bounds, with sensitivities competitive with
those of planned G2 direct detection experiments.

For antimatter, recent measurements of cosmic-ray positrons by the
AMS-02 magnetic spectrometer confirm and improve with excellent
precision earlier measurements by PAMELA and Fermi.  The rising
positron fraction could be indicative of positrons created in the
decay or annihilation of dark matter.  In the near future, AMS-02 will
extend its determination of the positron fraction to energies close to
1 TeV, and add important information on cosmic-ray propagation.  Given
the possibility of astrophysical sources of primary positrons,
however, it may be very difficult to definitively attribute the excess
positrons to dark matter.  Antideuterons provide a signal that is
potentially more easily discriminated from astrophysical
backgrounds. With a long-duration balloon flight, the General
Antiparticle Spectrometer (GAPS) experiment could provide
sensitivities comparable to AMS.  Last, the production of positrons
and electrons from dark matter annihilation also produces secondary
radiation.  Detection of signals with radio to X-ray frequencies has
the potential to probe the WIMP parameter space.

The Snowmass process also evaluated the prospects for non-WIMP
candidates, which could be some or all of the dark matter.  The axion
is particularly well-motivated, as it arises from the leading solution
to the strong CP problem of the Standard Model.  RF-cavity and solar
searches for axions, such as ADMX and IAXO, will probe a large range
of axion parameter space, including the cosmologically-favored region,
and have strong discovery potential.  Sterile neutrinos are also
highly motivated by the observed non-zero masses of active neutrinos.
In the mass range where sterile neutrinos are dark matter candidates,
their radiative decays produce a monoenergetic photon, which may be
detected with X-ray telescopes.  Many other dark matter candidates
were also surveyed, including asymmetric dark matter, primordial black
holes, Q-balls, self-interacting dark matter, superheavy dark matter,
and superWIMP dark matter.

How do the diverse strategies for identifying dark matter fit
together? The Snowmass process produced a clear articulation of how
the different approaches --- including the direct and indirect
detection experiments mentioned above, but also particle colliders and
astrophysical probes --- each provide unique and necessary
information.  This complementarity is discussed below and was examined
in two theoretical frameworks.  First, the discovery prospects were
examined in complete supersymmetric models, with randomly selected
parameters in the phenomenological MSSM framework.  Second, the
possibility that only the dark matter particle is kinematically
accessible was considered using the framework of dark matter effective
theories.  In both cases, the complementarity of different approaches
was evident at all levels, both to establish a compelling dark matter
signal and, just as importantly, after discovery, to determine the
detailed properties of the particle or particles that make up dark
matter.

\vspace*{0.05in} {\bf Dark energy and CMB}

Cosmic surveys --- optical imaging and spectroscopic surveys and
detailed measurements of the CMB --- precisely map the Universe on many
different angular scales and over wide ranges of cosmic time.  They
provide unique information about cosmology and new physics, including
inflation, dark matter, dark energy, and neutrino properties.  These
measurements are challenging, requiring advances in instrumentation
and excellent control of systematic effects.  Fortunately, these
advances are now within reach, thanks to decades of investment and
close collaborations between particle physicists and astrophysicists.
The payoffs for this effort are large.

Measurements of the distance--redshift relation, first using supernovae
and then additional complementary techniques, revealed the expansion
history of the Universe, particularly over the past several billion
years, and yielded the surprising discovery that the expansion rate
has been increasing instead of decreasing.  Now we must determine what
is causing the cosmic acceleration.  This ``dark energy'' must produce
negative pressure to be responsible for the observed effect.  One
important clue is whether the negative pressure has been constant in
time or is evolving.  The stage III (the DES and HSC imaging surveys,
and the PFS and eBOSS spectroscopic surveys) and stage IV (LSST
imaging survey and DESI spectroscopic survey on mountaintops; Euclid
and WFIRST-AFTA in space) dark energy facilities will constrain both
the value and the evolution of the value with much higher precision,
as recommended in previous community studies, but they will also do
much more.  We must also check whether our description of gravity is
correct, and this is where measurements of the growth of structure,
over a wide range of distance scales using both imaging and
spectroscopic surveys, are needed.

There are several alternatives to general relativity (GR) that can
accurately describe the observed distance--redshift relation, but they
also modify the behavior of gravity over different distance scales.
The alternative models therefore predict structure growth rates that
are different from those in the standard theory.  Measuring the
structure growth rate over many different distance scales will test GR
and the alternative models.  Deviation from expectation on just one of
these scales will signal new physics.  In other words, the upcoming
dark energy facilities, particularly at stage IV, where systematic
error management is built deeply into the design, will provide many
precise tests and will characterize the behavior of dark energy in
much richer ways than ``just'' the overall value and its evolution
with time.  We will know the strength of the effects in a
two-dimensional parameter space of distance and cosmic time, as well
as any deviations from expectations in the correlations among of them.
Further surprises may await us.

Inflation is the leading paradigm for the dynamics of the very early
Universe, and current observations of large-scale structure lend
support to this intriguing idea.  The most direct available probes of
inflation come from CMB observations, and the overall agreement is
remarkably good.  However, it has not been possible to explore the
underlying physics of inflation, until now: The coming generations of
CMB experiments will have sufficient sensitivity to falsify large
classes of models.  The signal is a characteristic pattern with
non-zero curl (called ``$B$ mode'') faintly imprinted on the
polarization of the CMB fluctuations, due to gravitational waves
produced during the epoch of inflation.  The shape of the potential of
the scalar field driving inflation directly affects the spectrum of
gravitational waves and hence the strength of the imprint, $r$ (the
ratio of tensor to scalar power), over characteristic angular scales
on the sky.  The current generation of experiments is sensitive to $r
\sim 0.1$, but over the next 10 to 20 years, improvements of two
orders of magnitude are possible by scaling the number of detectors by
similar factors, from $\sim 10^3$ (current) to $\sim 10^4$ (generation
III) to $\sim 5\times 10^5$ (generation IV).  This would require a
change from the way things have been done in the past.  Groups would
merge into one coordinated effort, and national lab facility design,
integration, computing, and management capabilities would be tapped.

Remarkably, future optical and CMB cosmic surveys, as well as future
polar-ice neutrino projects (see below), will also provide precise
information about neutrino properties, including the mass hierarchy,
the number of light neutrinos, and the sum of the neutrino species
masses.  Combining this information with accelerator- and
reactor-based neutrino experiments, as well as other experiments, such
as those searching for neutrinoless double-beta decay, will accelerate
our understanding of fundamental neutrino properties and enable us to
understand the implications of apparent inconsistencies.

Snowmass provided an excellent opportunity to address common problems
and to develop a common vision for the potential of cosmic surveys for
particle physics.  Highlights included advancing detailed strategies
to distinguish dark energy from modified gravity; exploiting the
complementarity of probes for determining the key cosmological
parameters; understanding more deeply the strengths and ultimate
limitations of the different techniques; and discussing the planned
facilities, which are the result of intensive community processes over
many years.  The group articulated a set of goals: (1) Remain a leader
in dark energy research, (2) build a generation IV CMB polarization
experiment, and (3) extend the reach of cosmic surveys with targeted
calibration campaigns, targeted R\&D, and support for work at the
interface of theory, simulation, and data analysis.

\vspace*{0.05in} {\bf Cosmic particles}

Measurements of fluxes of cosmic particles (charged particles,
photons, and neutrinos) address many topics in particle physics beyond
indirect dark matter searches.  Recent results include the detection
by IceCube of very high energy neutrinos that are likely to be from
astrophysical sources; the observation of the GZK suppression in the
cosmic-ray flux above $3\times 10^{19}$ eV; the measurement of the
positron fraction up to 300 GeV, suggesting the existence of primary
sources of positrons from astrophysical processes and/or dark matter
interactions; and confirmation that supernova remnant systems are a
source of galactic cosmic rays.  These and other discoveries were made
by the current generation of experiments.

Goals for the coming decade include determination of the origin of the
highest energy particles in the Universe, measuring interaction
cross sections at energies unattainable in terrestrial accelerators,
detecting the GZK neutrinos that arise from the interactions of
ultra-high-energy cosmic rays with the CMB, determining the neutrino
mass hierarchy, and searching for other physics beyond the Standard
Model.

To meet these goals, the group recommends: significant
U.S. participation in the Cherenkov Telescope Array (CTA), which is
the next-generation ground-based gamma-ray facility; simultaneous
operation of Fermi, HAWC, and VERITAS, the current generation of
U.S.-led space- and ground-based gamma-ray facilities; construction of
the PINGU neutrino detector to lower the energy threshold to a few GeV
and enable the determination of the neutrino mass hierarchy using
atmospheric neutrinos; continued operation of the Auger and Telescope
Array air shower arrays with upgrades to enhance the determination of
the composition of the flux of cosmic rays around the GZK suppression
region; construction and deployment of the JEM-EUSO mission aboard the
International Space Station to extend observations of the cosmic ray
flux and anisotropy well beyond the GZK region; and construction of a
next-generation ultra-high-energy GZK neutrino detector, which will
either detect GZK neutrinos (and constrain the neutrino-nucleon cross
section at ultra-high energy) or exclude all but the most unfavorable
parts of the allowed parameter space.  A detailed census of present
and proposed cosmic particle measurement facilities was produced
during the Snowmass process.

\vspace*{0.05in} {\bf Conclusions}

In synergy with the other Frontier areas, the Cosmic
Frontier provides particle physics clear evidence for physics
beyond the Standard Model; profound questions of popular interest;
frequent new results; surprises, with broad impacts; a large discovery
space with unique probes; important cross-frontier topics; and a full
range of project scales, providing flexible programmatic options.  For
each area of the Cosmic Frontier, the Snowmass study identified
essential technologies and facilities, the advances required in
theoretical models, and experiments with great promise.  The largest
projects are, appropriately and necessarily, international.  The
U.S. is still the leader in many areas of the Cosmic Frontier, but
this field is evolving quickly and other regions with intensive
interest in this physics are advancing rapidly.

\section{Direct detection of WIMP dark matter}
\label{sec:direct}

Deciphering the nature of dark matter is one of the primary goals of
particle physics for the next decade. Astronomical evidence of many
types, including cosmic microwave background (CMB) measurements,
cluster and galaxy rotation curves, lensing studies and spectacular
observations of galaxy cluster collisions, all points to the existence
of CDM particles. Cosmological simulations based on
the CDM model have been remarkably successful at predicting the actual
structures we see in the Universe~\cite{Navarro:1996gj}. Alternative
explanations involving modification of Einstein's theory of GR have
not been able to explain this large body of evidence across all length
scales.

WIMPs are strong candidates to explain dark matter.  They represent a
class of dark matter particles that froze out of thermal equilibrium
in the early Universe with a relic density that matches observation.
This coincidence of scales --- the relic density and the weak force
interaction scale --- provides a compelling rationale for WIMPs as
particle dark matter. There are many particle physics theories that
provide natural candidates for WIMPs, but they do not limit the search
parameters very much, leaving a search region with a range of 1 GeV to
100 TeV in mass and $10^{-40}$ to $10^{-50}$~cm$^2$ in interaction
cross sections with normal matter.

Direct detection experiments are designed to identify the interaction
of WIMPs with normal matter. Since WIMPs should interact with normal
matter by elastic scattering with nuclei~\cite{Goodman:1984dc}, this
requires detecting nuclear recoil energies in the 1--100 keV
range. These low energies and cross sections represent an enormous
experimental challenge, especially in the face of daunting backgrounds
from electron recoil interactions and from neutrons that mimic the
nuclear recoil signature of WIMPs. The unambiguous detection of signal
above these backgrounds would confirm the existence of WIMPs in our
galaxy and begin to unravel the mystery of their identity, especially
with complementary information from production in colliders and
signals from annihilation in our galaxy or in the Sun.

Direct detection experiments must be located in deep underground
laboratories to avoid effects of cosmic-ray interactions that produce
energetic neutrons that could mimic WIMPs.  The experiments must also
shield the detectors from the decay products of radioactivity in the
environment and in the materials of the experiment itself.  This is
especially important for neutrons resulting from fission or $(\alpha,
n)$ reactions, since a single scatter from a neutron produces a
nuclear recoil that is indistinguishable from that produced by a WIMP.
Luckily, some portion of the neutrons will scatter multiple times in
the detector, and so neutron events can be identified and rejected on
this basis.  In most cases, the experiments also define an ultra-pure
active ``fiducial'' volume that is shielded from radioactive decay
products produced by impurities in the materials surrounding the
detection material.

\begin{figure}[tb]
\begin{center}
\includegraphics[width=.83\hsize]{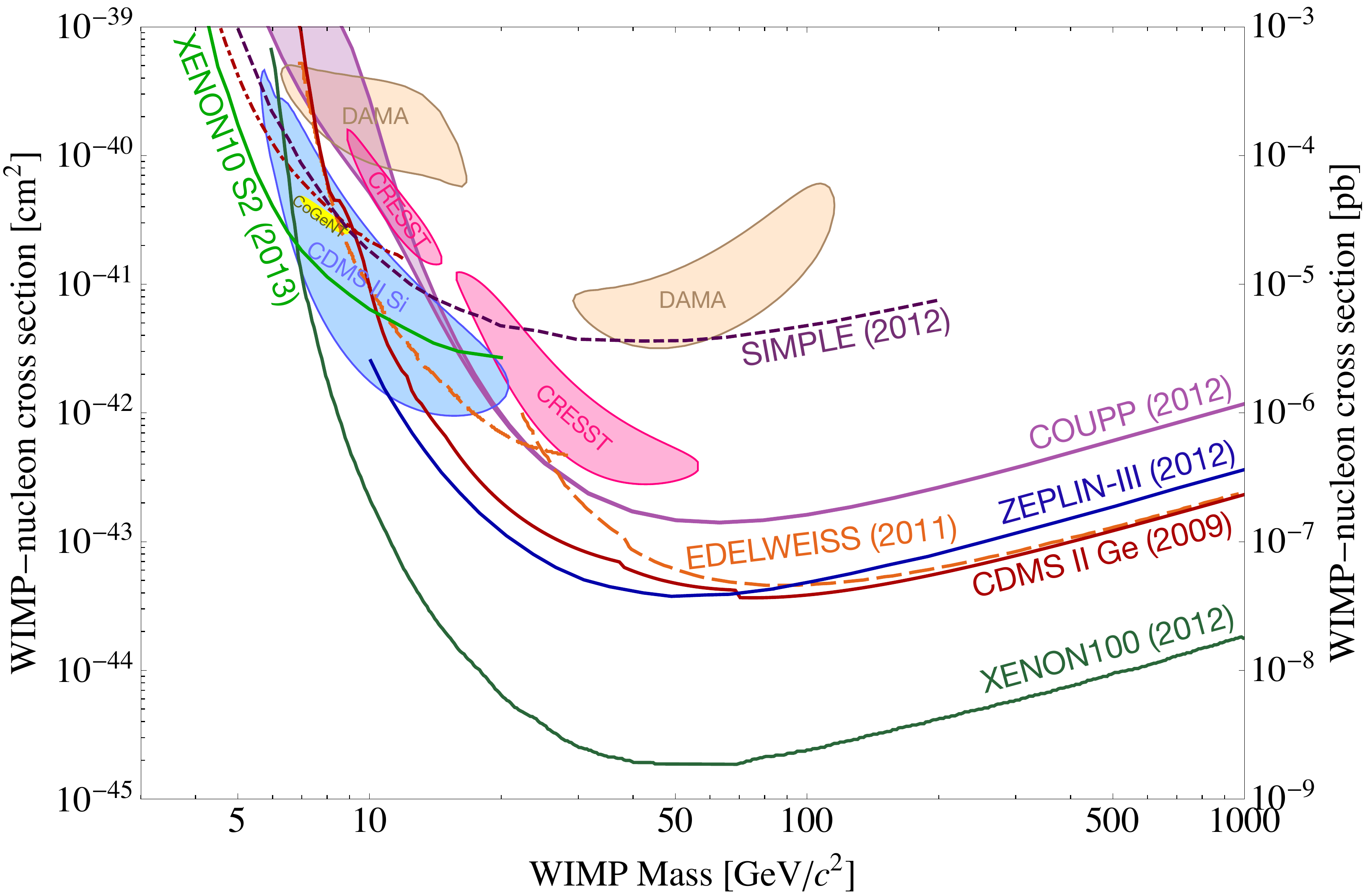}
\vspace*{.1in}
\caption{Constraints on spin-independent WIMP-nucleon cross sections
  as a function of WIMP mass as of Summer 2013~\cite{Angle:2011th,%
    Aprile:2012nq,Aprile:2011hi,Girard:2012zz,Behnke:2012ys,%
    Aalseth:2012if,Angloher:2011uu,Bernabei:2010mq,Armengaud:2011cy,%
    Akimov:2011tj,Agnese:2013rvf,Agnese:2013cvt,Ahmed:2009zw,%
    Ahmed:2010wy}.
\label{fig:fig001} }
\end{center}
\end{figure}

\begin{figure}[tbp]
\begin{center}
\includegraphics[width=.73\hsize]{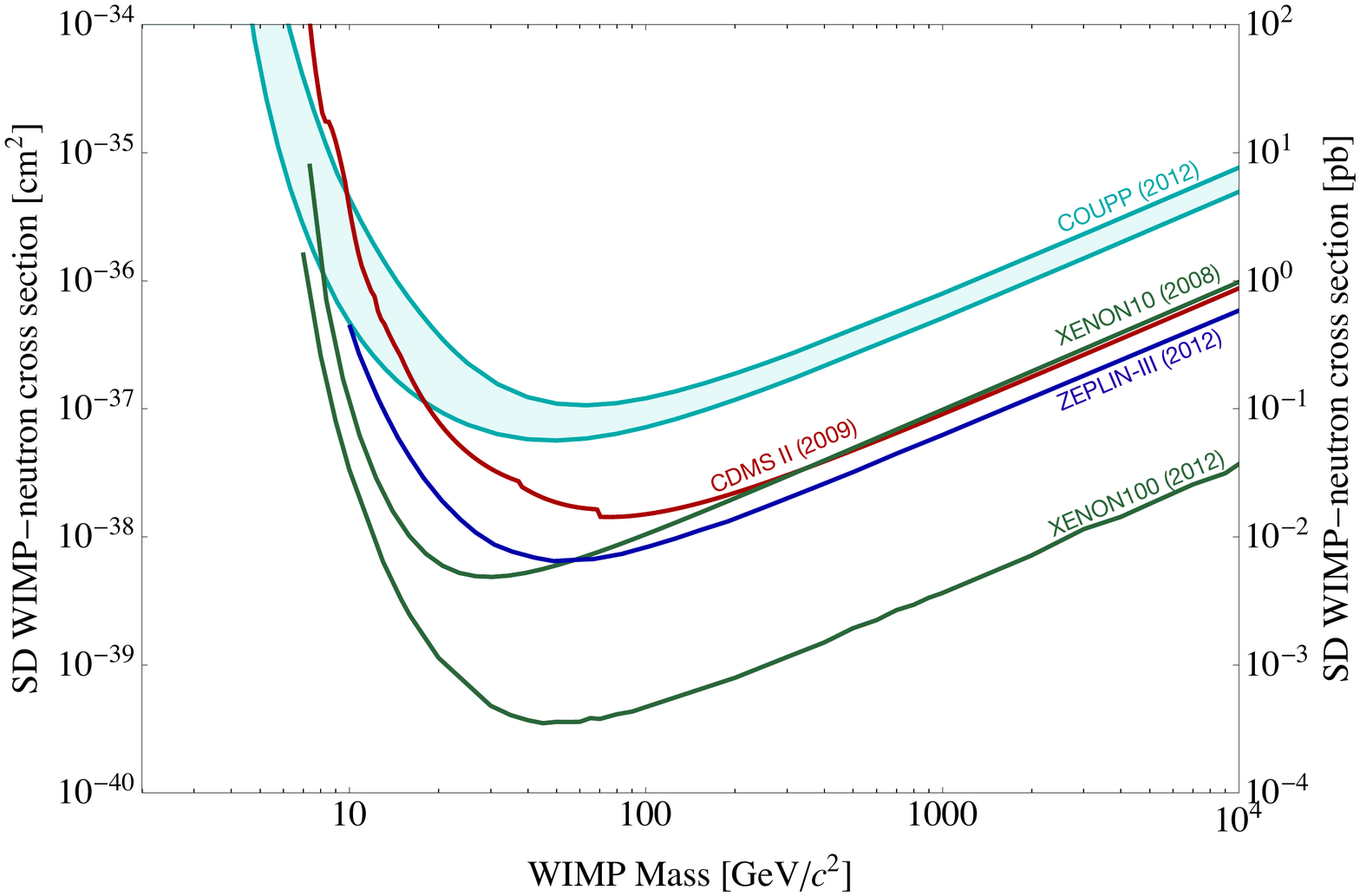}
\includegraphics[width=.73\hsize]{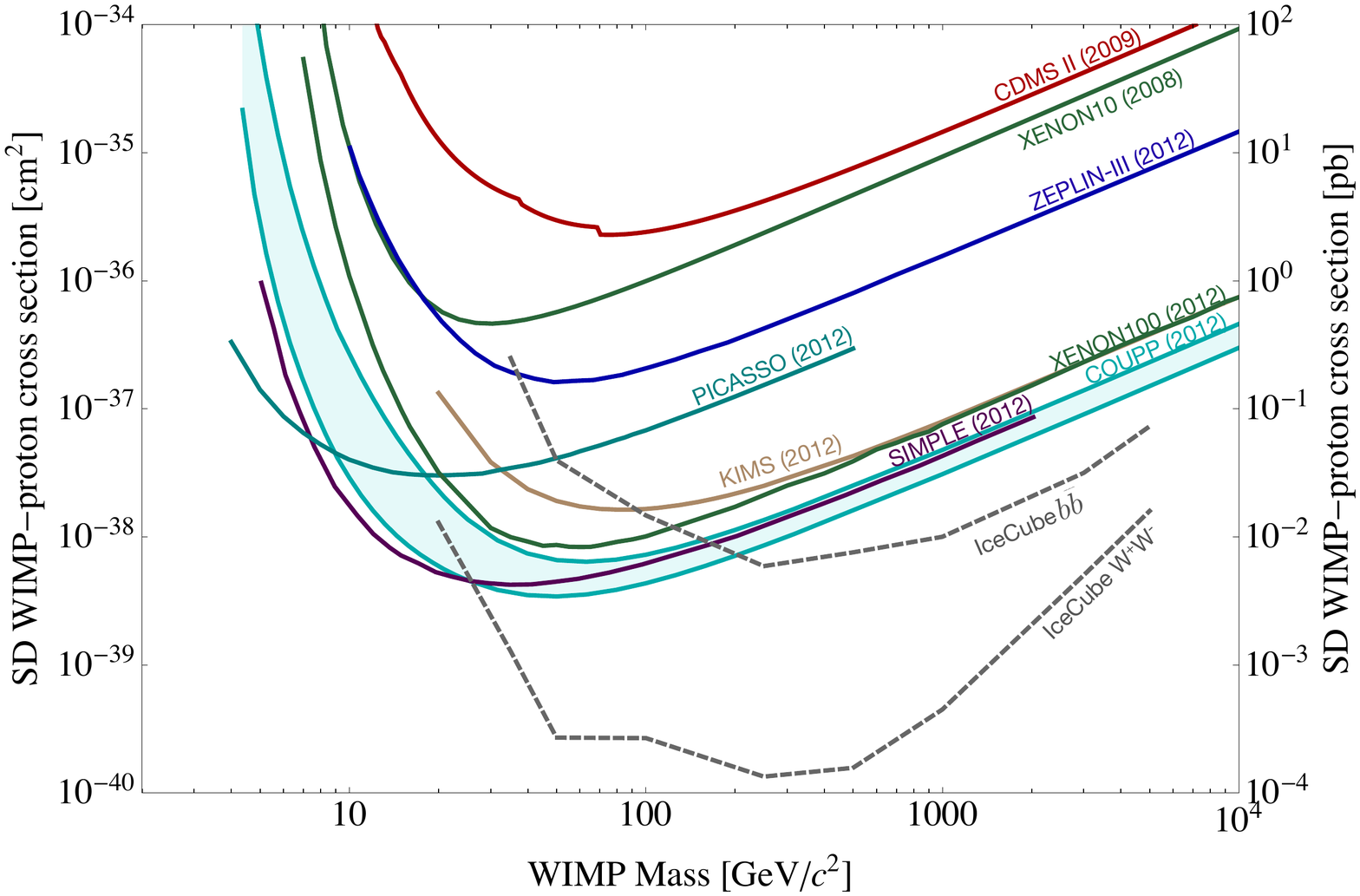}
\vspace*{-.2in}
\caption{Spin-dependent WIMP-neutron (left) and WIMP-proton (right)
  cross section limits as functions of WIMP mass for direct detection
  experiments~\cite{Aprile:2013doa,Akerib:2005za,Akimov:2011tj,%
    Archambault:2012pm,Kim:2012rza,Behnke:2012ys,Girard:2012zz}, as
  well as IceCube results (model-dependent)~\cite{Aartsen:2012kia}, as
  of Summer 2013. }
\label{fig:fig0023}
\end{center}
\end{figure}

The basic methodology for direct detection experiments is to search
for elastic scattering of a WIMP from a target nucleus.  The rate of
candidate nuclear recoils is converted into a cross section for
WIMP-nucleon interactions following a standard prescription that
includes the effects of nuclear physics and astrophysical
properties~\cite{Lewin:1995rx}.  Experiments can be sensitive to both
nuclear spin-independent (SI) interactions and spin-dependent (SD)
interactions.  For the momentum exchange range of interest, the SI
interaction is expected to be approximately coherent across the entire
nucleus, so for a WIMP with equal coupling to protons and neutrons,
the rate scales with the square of the atomic mass of the target
nucleus. Current experiments are therefore more sensitive to SI dark
matter than SD dark matter.  Experimental results are usually
presented as a plot of WIMP-nucleon cross section versus WIMP mass to
allow comparison among experiments. \Figref{fig001} shows the current
SI landscape, with strict upper limits for higher mass WIMPs and some
closed contours for lower mass WIMPs. The SD interaction is generally
divided into proton and neutron couplings; the results to date are
summarized in \Figref{fig0023}. Only direct detection can provide
limits on neutron couplings, but solar neutrinos from WIMPs that
annihilate in the Sun are stronger for proton coupling.  Other types
of interactions are possible, and it is important to run multiple
experiments with different targets, both to cover the parameter space
for discovery and to study the interaction type when signals are
found.

Nuclear recoils from WIMP scattering result in a featureless energy
spectrum, rising exponentially as the energy decreases. Experiments
typically do not directly measure the nuclear recoil energy. Instead,
the energy deposited by a particle interaction must be reconstructed
from the experimental measurements as either nuclear-recoil (keV$_r$)
or electron-recoil (keV$_{ee}$).  Conversion between the two energies
is dependent on the target and experimental technique, and must be
calibrated by each experiment. Typically, radioactive gamma sources or
in-situ doped beta emitters are used to provide calibration for
electron recoils, and neutron sources provide a source of nuclear
recoil events.  Since the nuclear recoil calibration is difficult at
low energies, there is significant uncertainty in the low WIMP mass
exclusion limits, sowing some controversy when comparing limits and
discovery contours between targets.

A complete list of the current targets and technologies in use for
direct detection of WIMPs can be found in the CF1
Summary~\cite{Cushman:2013zza}. One of the key features of modern
experiments is the use of discrimination to select nuclear recoils and
reject backgrounds that are dominated by electron recoils. Often the
signal is split into two components that respond differently to
nuclear and electron recoils. Noble liquids, such as argon and xenon,
make use of ionization and scintillation light. Solid targets, such as
silicon and germanium at cryogenic temperatures, use heat (or phonons)
plus ionization.  Crystals, such as CaWO$_{4}$, contrast phonons and
scintillation light.  Two components are not necessary if an
experiment can achieve internal purities and excellent fiducial
isolation, or can use pulse-shape discrimination (\eg, liquid argon),
or is insensitive to electron recoil backgrounds (\eg, threshold
detectors).

Another method to deal with backgrounds is to exploit the fact that
the Earth is moving through the dark matter that surrounds our galaxy,
yielding a ``WIMP wind'' that appears to come from the constellation
Cygnus~\cite{Drukier:1986tm}.  This should, in principle, create a
small ``annual modulation'' in the detected WIMP rates, as well as a
somewhat larger daily modulation. However, if such effects were
detected in an experiment, there would still have to be a convincing
demonstration that there are no such modulations in background
sources.  Since backgrounds are expected to be isotropic, detection of
a signal with a preferred direction could provide a powerful
additional discriminant against backgrounds. Directional detectors
attempt to exploit this effect by sensing the vector direction of
nuclear recoils, and they are usually looking for a daily sidereal
difference as the Earth rotates relative to the WIMP wind.

\begin{figure}[tb]
\begin{center}
\includegraphics[width=.83\hsize]{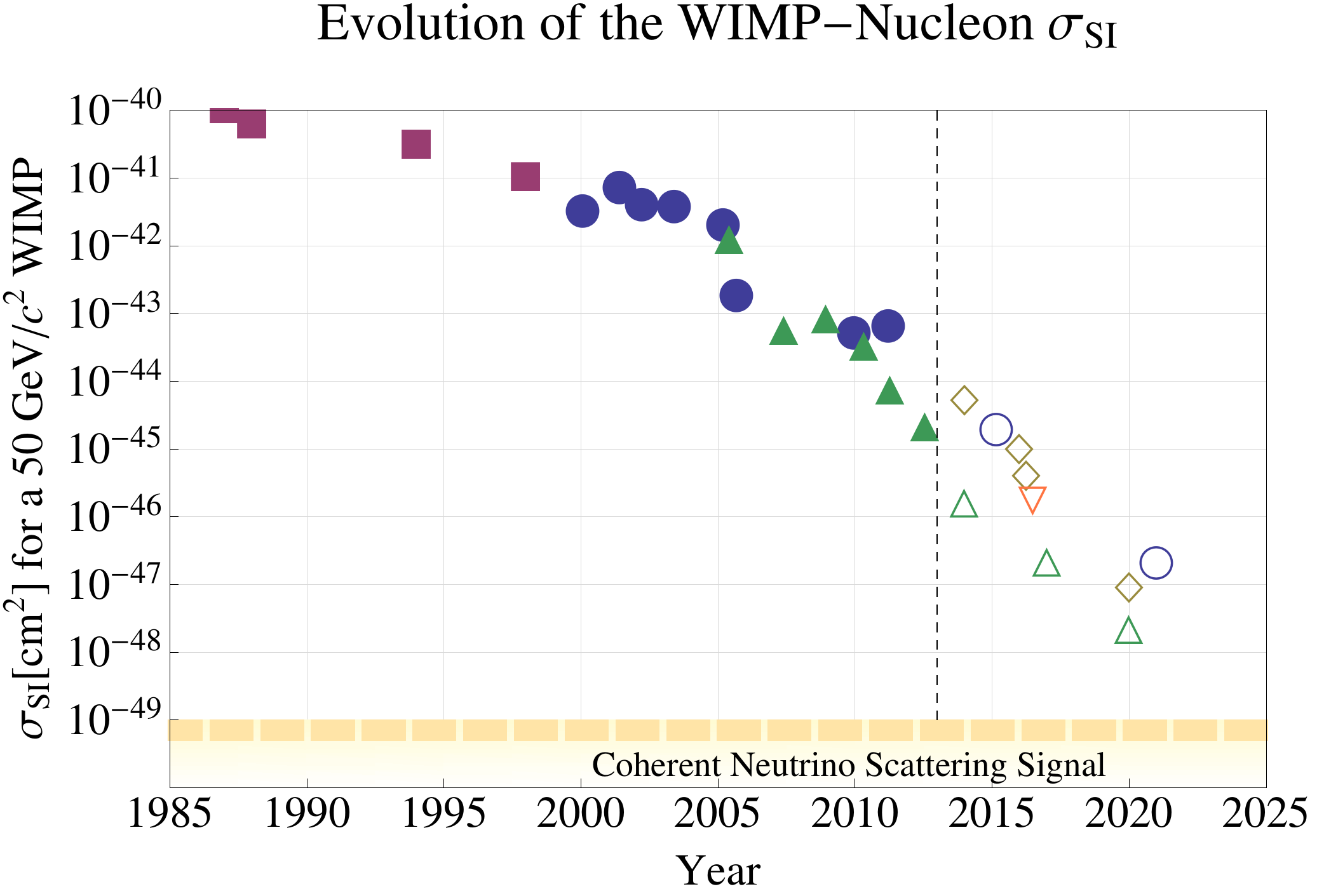}
\vspace*{0.1in}
\caption{Spin-independent limits for the major WIMP direct detection
  experiments (solid) and their projected sensitivity (open) for spin
  independent cross sections for a 50 GeV/$c^2$ mass WIMP. The shapes
  correspond to technologies: cryogenic solid state (blue circles),
  crystal detectors (purple squares), liquid argon (brown diamonds),
  liquid xenon (green triangles) and threshold detectors (orange
  inverted triangle). }
\label{fig:fig004}
\end{center}
\end{figure}

The first two decades of dark matter direct detection experiments have
yielded a diverse and successful program, although not yet definitive
evidence for WIMPs.  Starting with just a few experiments using
solid-state targets, the technologies used for these experiments have
grown considerably.  There has been a remarkable improvement in WIMP
sensitivities, especially in the range where the WIMP mass is
comparable to the atomic mass of the target nuclei.  A selection of
spin-independent results from the first two decades of these
experiments, and projections for the coming decade, is shown in
\Figref{fig004} for a 50 GeV/$c^2$ WIMP mass.  A ``Moore's law''-type
improvement is particularly evident for such WIMP masses, with a
sensitivity doubling time of roughly 18 months.  Note that direct
detection experiments have sensitivity to much larger WIMP masses as
well, surpassing what is accessible at the LHC. More recently, there
has also been rapid progress in sensitivity to WIMPs with masses at 10
GeV/$c^2$ and below.

\begin{figure}[tb]
\begin{center}
\includegraphics[width=.93\hsize]{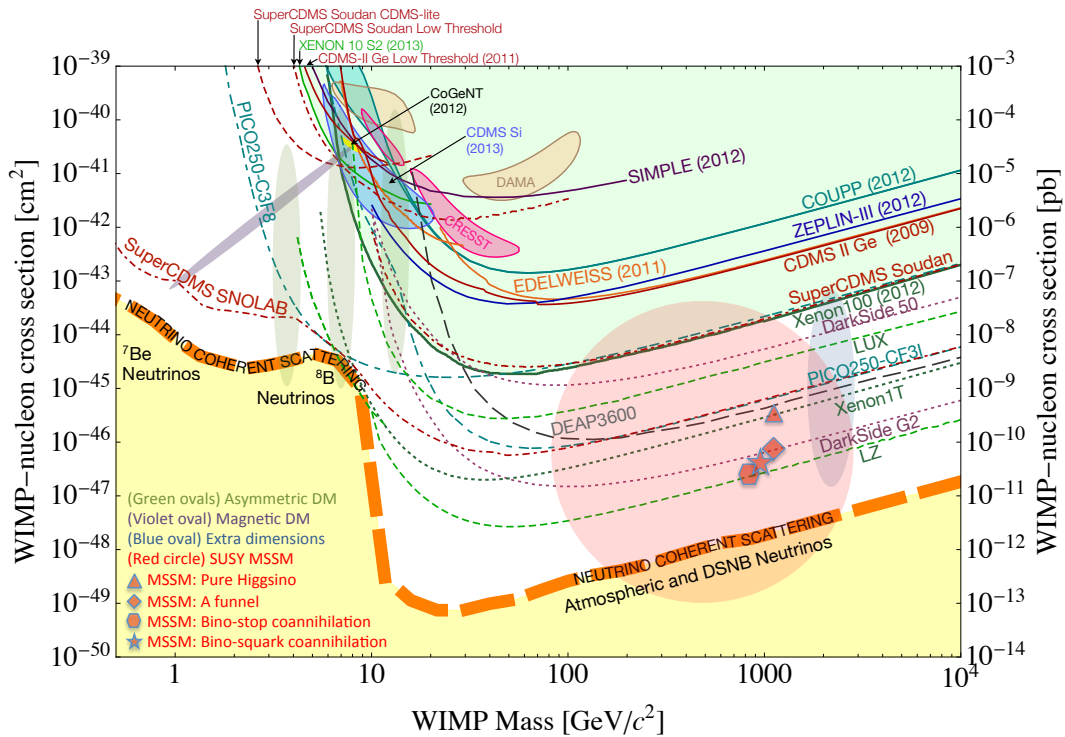}
\vspace*{-.2in}
\caption{A compilation of WIMP-nucleon spin-independent cross section
  limits (solid curves), hints for WIMP signals (shaded closed
  contours) and projections (dot and dot-dashed curves) for U.S.-led
  direct detection experiments that are expected to operate over the
  next decade. Also shown is a band indicating the cross sections
  where WIMP experiments will be sensitive to backgrounds from solar,
  atmospheric, and diffuse supernovae neutrinos. }
\label{fig:fig005}
\end{center}
\end{figure}

Sensitivity projections are subject to uncertainties from many
factors, including technical issues with the experiments, the
appearance of unexpected backgrounds, and delays in funding. Despite
these uncertainties, the history of the field gives us confidence that
progress will continue unabated through the next decade.  There is,
however, an irreducible background caused by solar, atmospheric, and
supernova neutrinos.  The $^8$B component of the solar neutrinos
provides the dominant neutrino coherent scattering rate at low recoil
energies. The $^8$B spectrum can mimic a WIMP with a mass in the range
of 5--10 GeV/$c^2$ (depending on the target) and a nucleon cross
section of $\sim 5 \times 10^{-45}$~cm$^2$. For experiments targeted
at larger WIMP masses, the solar neutrinos give way to the more
energetic atmospheric neutrinos and diffuse supernovae background. The
flux of these neutrinos is much lower, and exposures with
sensitivities to WIMP-nucleon cross sections of $\sim 1 \times
10^{-48}$~cm$^2$ are required to be sensitive to this neutrino
component.  Depending on the particular WIMP mass under consideration,
these neutrino backgrounds can have a recoil spectrum that is very
similar to an authentic WIMP signal. Given the Poisson fluctuations
from the neutrino signal and their relatively large total flux
uncertainties, this creates a challenge to improving the sensitivity
of WIMP searches much beyond such cross
sections~\cite{Billard:2013qya}. \Figref{fig005} shows not only the
current landscape, but also the projected sensitivities of proposed
experiments superimposed on the neutrino background, where coherent
neutrino scattering will begin to limit WIMP sensitivity.  This will
eventually require either background subtraction or techniques such as
directional or annual modulation to press beyond this background in
the absence of a positive WIMP sighting.

\section{Indirect detection of WIMP dark matter}
\label{sec:indirect}

If thermal decoupling is the mechanism that sets the abundance of WIMP
dark matter in the early Universe, WIMPs generically are predicted to
annihilate today as well, especially in regions of high dark matter
density, producing Standard Model particles, including gamma rays,
neutrinos, electrons, positrons, protons, antiprotons, deuterons, and
antideuterons.  Utilizing these Standard Model ``messengers'' to
constrain or infer properties of the dark matter is the ultimate goal
of indirect detection.

The dark matter annihilation signal depends both on particle physics
properties, such as the annihilation cross section, and on the density
distribution of dark matter. The latter can be determined either by
$N$-body simulations of structure formation in the $\Lambda$CDM
standard model of cosmology or by direct dynamical measurements that
derive the enclosed mass (luminous and dark) by measuring the velocity
dispersion of stars or molecular clouds.  In the absence of baryons,
CDM simulations indicate that dark matter halos are cuspy (with
densities approximately scaling as $\rho \sim r^{-1}$ in their
centers~\cite{Navarro:1996gj,navarro2004,navarro2010}), with a central
concentration dependent on the mass of the halo and the halo formation
time~\cite{Bullock:1999he,wechsler2002}.  Small halos tend to form
earlier than big halos and also to be much more densely concentrated.
However, the places where we have the highest dark matter annihilation
rates (\ie, the centers of halos) are also the places where baryons
settle and dominate the potential well for typical halos.  There is
not yet a consensus on how baryons affect dark matter
halos~\cite{blumenthal1984,gnedin2004,tissera2010,martizzi:2012ci,%
  governato2012}.  Fortunately, even when one makes the most
conservative assumptions about the halo profiles (\eg, cored profiles
constrained by dynamics), indirect detection methods still constrain
the viable WIMP parameter space.

\subsection{Gamma-ray experiments}

The ability to detect the dark matter signal from a given target
depends critically on its dark matter density distribution and on $J$,
the integral of the square of the dark matter density along the line
of sight to the source. The ideal targets for dark matter annihilation
searches are those that have both a large value of $J$ and relatively
low astrophysical $\gamma$-ray foregrounds. These criteria have
motivated a number of galactic and extragalactic targets including the
galactic center (GC), dwarf spheroidal satellite galaxies of the Milky
Way (dSphs), and galaxy clusters.

\begin{figure}[tb]
\begin{center}
      \includegraphics[width=0.83\hsize]{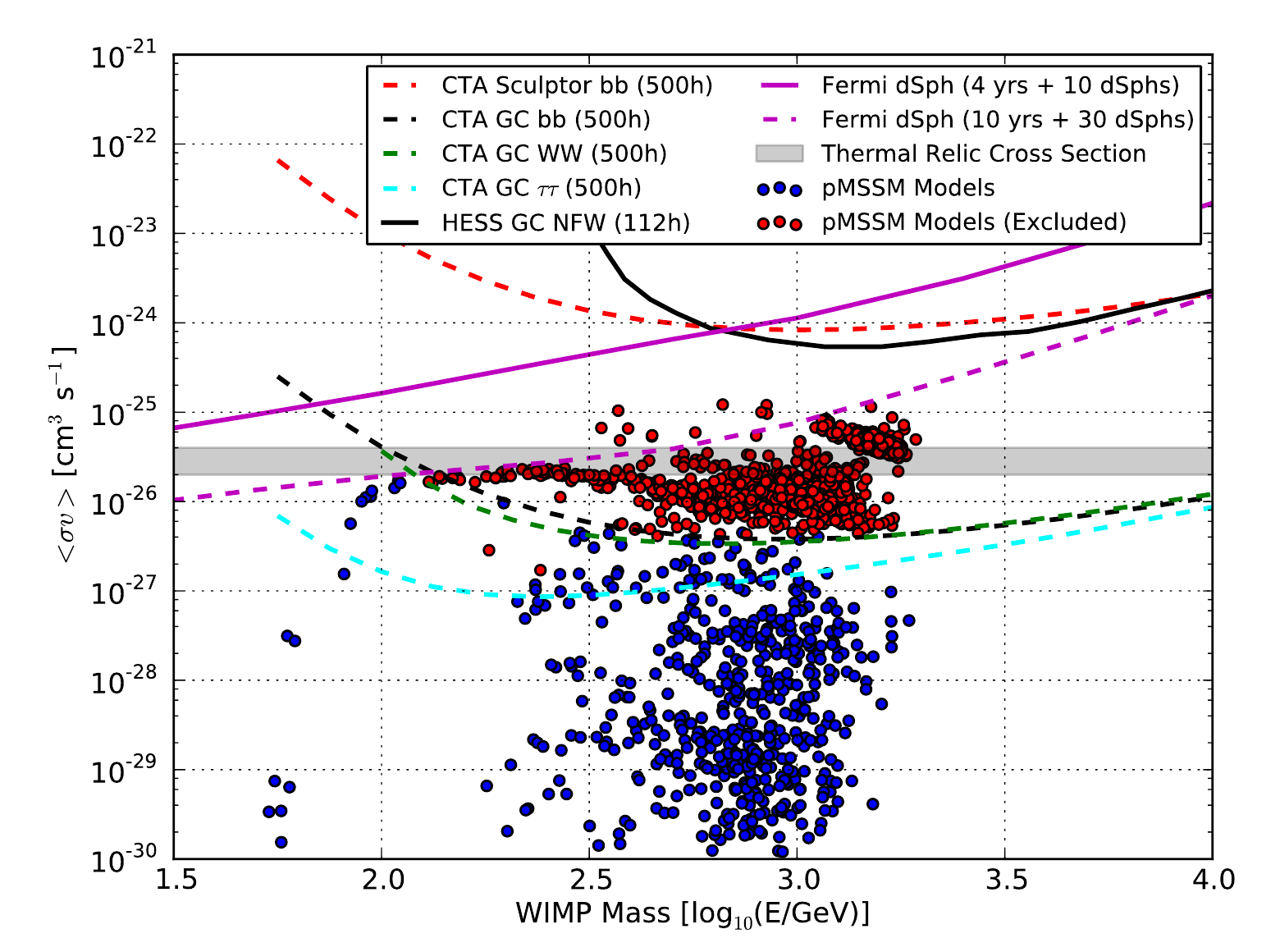}
\vspace*{.1in}
\caption{Constraints on the total annihilation cross section
  from $\gamma$-ray experiments, including the anticipated sensitivity
  of Fermi obtained by stacking 10 yrs of data on dSphs, and the
  sensitivity of the augmented CTA instrument, using an annulus about
  the GC. }
   \label{fig:ctasens1}
\end{center}
\end{figure}

In and near the GC, the dark matter-induced $\gamma$-ray emission is
expected to be so bright that one can obtain strong upper limits at
the level of the target annihilation cross section $\langle
\sigma_{\rm{an}} v \rangle \sim 3 \times 10^{-26}~{\rm cm}^3~{\rm
  s}^{-1}$, even after excluding regions around the bright source at
the GC and surrounding region.  With the improved angular resolution
of future $\gamma$-ray experiments, the astrophysical foregrounds will
be more easily identified and separated from the diffuse annihilation
signal, and if, \eg, the GC source is a point source, the dark matter
sensitivity of future experiments could exceed the predictions
presented here (see \Figref{ctasens1}).

\begin{figure}[tb]
\begin{center}
      \includegraphics[width=0.83\hsize]{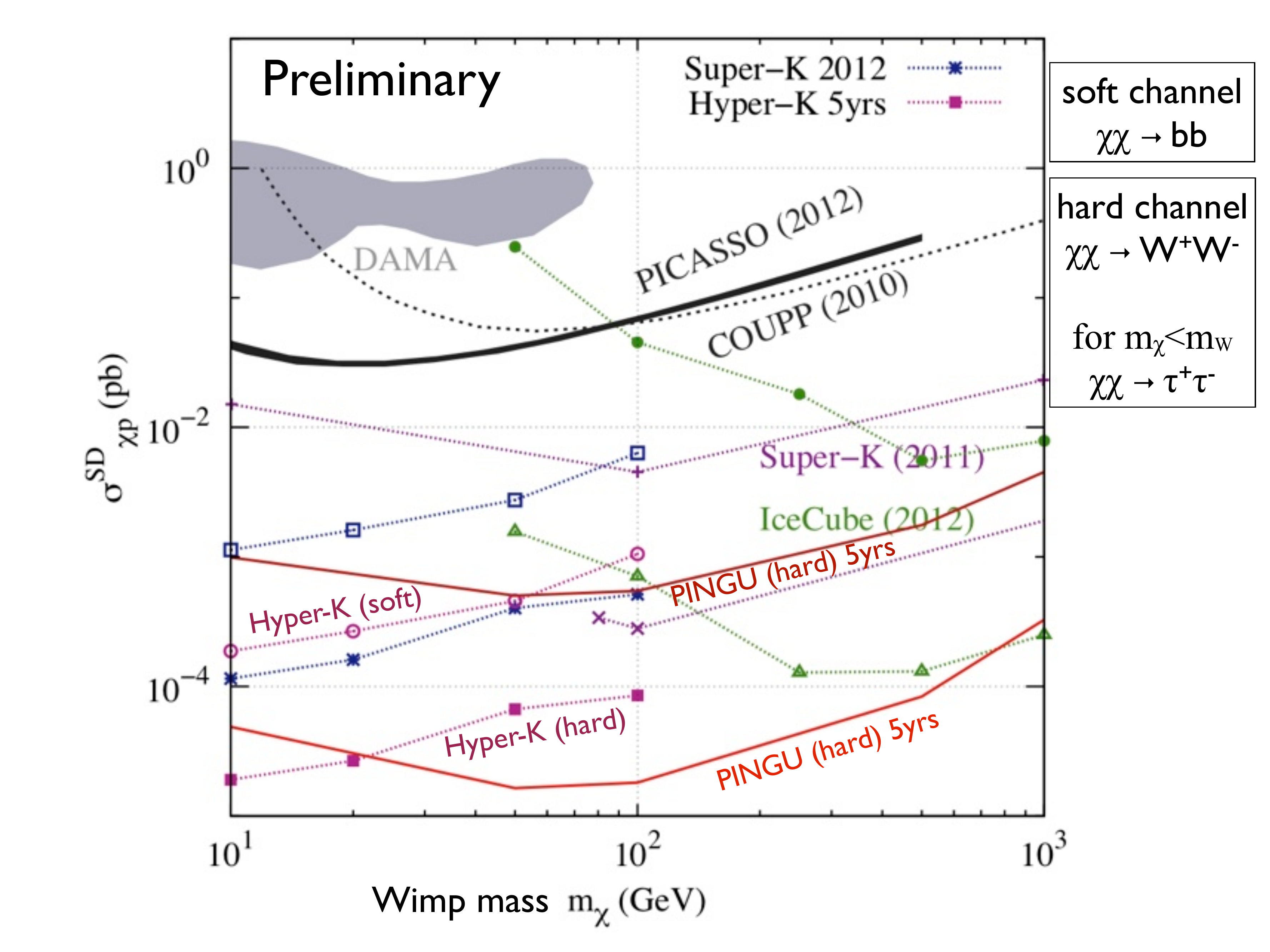}
\vspace*{.1in}
\caption{Existing constraints on the
  spin-dependent WIMP-proton scattering cross section from various
  experiments, along with projected future sensitivities for
  Hyper-Kamiokande and PINGU with five years of data.  }
   \label{fig:ctasens2}
\end{center}
\end{figure}

The Milky Way is also known to host at least two dozen dSphs that are
typically very dark-matter-dominated, with mass-to-light ratios of 100
to 1000, and are thought to have very small background emission of
high energy $\gamma$-rays.  About half of the dSphs were discovered in
the Sloan Digital Sky Survey (SDSS).  A large number of additional
dSphs are believed to exist in the Milky Way
halo~\cite{koposov2008,tollerud2008}.  Big, deep, wide-field surveys
like the Dark Energy Survey and, critically, LSST, should find more.
In particular, these experiments should help to find additional nearby
galaxies in the southern hemisphere, since SDSS did not explore this
area.  These objects are of particular importance to the future
southern CTA experiment that is described below.

The Fermi-LAT data already exclude the target annihilation cross
section for masses below 30 GeV, but only for the most optimistic
phenomenological minimal supersymmetric standard model (pMSSM) models.
In the low-energy, background-dominated regime, the Fermi-LAT point
source sensitivity increases roughly as the square root of the
integration time.  However, in the high-energy, limited-background
regime (where many pMSSM models predict signals), the Fermi-LAT
sensitivity increases more linearly with integration time.  Thus, 10
years of data could provide a factor of $\sqrt{5}$ to 5 increase in
sensitivity.

VERITAS has produced some of the best dark matter constraints at
energies above a few hundred GeV through observations of one of the
nearest, most dark-matter-dominated dwarf galaxies, Segue
1~\cite{Aliu:2012ga}.  This analysis was based on about 40 hours of
data, but the current VERITAS program, augmented by recent upgrades,
aims to observe dark matter targets for several hundred hours each
season.  This offers one of the strongest avenues available for WIMP
dark matter detection and, in the absence of discovery, can provide
severe constraints on many dark matter models.

For the next generation experiment, improving dark matter searches in
the GC depends on stronger astrophysical background rejection through
improved angular resolution.  CTA~\cite{Acharya:2013sxa} is a future
ground-based $\gamma$-ray observatory that will have potentially
game-changing sensitivity over the energy range from a few tens of GeV
to a few hundreds of TeV.  Doubling the size of the proposed southern
CTA telescope, a device with the required resolution and good
sensitivity to the GC and southern sources, will dramatically improve
dark matter sensitivity.  To achieve the best sensitivity over this
wide energy range, CTA will include three distinct telescope sizes.
Over this energy range, the point-source sensitivity of CTA will be at
least one order of magnitude better than current generation imaging
atmospheric Cherenkov telescopes (IACTs).  CTA will also have an
angular resolution at least 2 to 3 times better.  In the following
sensitivity estimates, we consider an augmented version of the CTA
array (ACTA) with a U.S. enhancement, which has twice as many
medium-sized telescopes ($\sim 60$ in total) and triple the
sensitivity.

\Figref{ctasens2} shows the projected sensitivity of ACTA to a WIMP
particle annihilating through three possible final states: $b\bar{b}$,
$W^+W^-$, and $\tau^+\tau^-$.  For an observation of the GC utilizing
a 0.3$^\circ$--1.0$^\circ$ annular search region, ACTA could exclude
models with cross sections significantly below the target annihilation
cross section.  Overlaid in the figure are WIMP models generated in
the pMSSM framework that satisfy all current experimental constraints
from collider and direct detection searches.  Approximately half of
the models in this set could be excluded at the 95\% confidence level
in a 500-hour observation of the GC.

ACTA, with the critical U.S. enhancement, will provide a powerful new
tool for searching for dark matter, covering parameter space not
accessible to other techniques. ACTA will provide new information to
help identify the particle nature of the dark matter and determine the
halo profile.  With support levels comparable to the complementary
searches of the G2 direct detection experiments, the U.S. dark matter
program would have the realistic prospect of both detecting dark
matter in the lab and identifying it in the sky.

The HAWC detector will complement existing IACTs and the space-based
$\gamma$-ray telescopes with its high-energy sensitivity and its large
field-of-view. The instrument has peak sensitivity to annihilation
photons from WIMP dark matter with masses between about 10 TeV and the
unitarity limit $\sim$100 TeV.  Such large masses can still satisfy
all cosmological and particle physics constraints for a generic
WIMP~\cite{profumo05}.  Much like Fermi, HAWC will survey the entire
northern sky with sensitivity roughly comparable to existing IACTs and
will search for annihilation from candidates that are not known {\em a
  priori}, such as galactic substructure. Furthermore, HAWC can search
for sources of gamma rays that are extended by 10 degrees or more and
can constrain even nearby sub-halos of dark matter.

\subsection{Charged cosmic-ray/antimatter experiments}

Dark matter may annihilate through a number of channels (\eg,
$t\bar{t}$, $b\bar{b}$, $W^+W^-$, $ZZ$, $\tau^+ \tau^-$) with similar
branching ratios.  These annihilation channels can eventually lead to
cascades producing secondary particles, such as $e^\pm$, $\gamma$, and
$\nu$.  Annihilation can also lead to the production of baryons, such
as cosmic-ray protons and antiprotons and even deuterons and
antideuterons.

Cosmic-ray electrons and positrons provide a unique astrophysical
window into our local galaxy.  To a good approximation, the magnetic
fields in the galaxy randomize cosmic-ray directions.  However, a
small anisotropy may remain due to contributions from local sources,
such as dark matter subhalos or nearby pulsars.  The identification of
a dark matter signal from cosmic rays thus requires the detection of a
spectral feature that stands out against the background.

Recent measurement of cosmic-ray positrons by the AMS-02 magnetic
spectrometer~\cite{Aguilar:2013qda} confirms with excellent precision
earlier measurements by PAMELA~\cite{Adriani:2008zr} and
Fermi~\cite{FermiLAT:2011ab} showing a rising positron fraction,
$e^+/(e^+ + e^-)$, for energies between 10 and several 100
GeV~\cite{FermiLAT:2011ab,Adriani:2008zr}.  The most widely used
models of propagation of cosmic rays in the galaxy~\cite{ms1998}
predict that secondary positrons would give a positron fraction
falling well below the observed value.  The rising positron fraction
could be indicative of positrons created in the decay or annihilation
of dark matter~\cite{2009PhRvL.103c1103B,Hektor:2013yga}.  Under this
interpretation, the dark matter particles must have mass greater than
350 GeV.  AMS-02 is capable of identifying cosmic-ray positrons with
energies up to $\sim 1$ TeV.  With several more years of data, AMS-02
should be able to extend its determination of the positron fraction to
energies close to 1 TeV and add important information on cosmic-ray
propagation. However, the possibility of astrophysical sources of
primary positrons remains a source of great concern and makes it
difficult to clearly attribute the excess positrons to dark matter
annihilation~\cite{2009JCAP...01..025H,2013PhRvD..88b3013C}.

About a decade ago it was pointed out that antideuterons produced in
WIMP-WIMP annihilations offered a potentially attractive signature for
CDM~\cite{donato00}.  The General Antiparticle Spectrometer (GAPS)
detector~\cite{hailey09} would consist of a detector that identifies
antideuterons using a number of planes of Si(Li) solid state detectors
and a surrounding time-of-flight system.  A long-duration balloon
flight of the proposed GAPs instrument could provide upper limits
competitive with those obtained by AMS-02, and would provide two
independent approaches to making this important measurement.

\subsection{Neutrino measurements}

Like their $\gamma$-ray counterparts, neutrino telescopes can
potentially detect the products of WIMP annihilations pointing back to
their origin in the GC, the galactic halo, from galaxy clusters, and
from dSphs.  Neutrino telescopes are also sensitive to WIMP
annihilations in the core of the Earth or Sun, regions that are
inaccessible to $\gamma$-ray telescopes.  The WIMP source in the Sun
has built up over solar time, averaging over the galactic dark matter
distribution as those WIMPs that scattered elastically with solar
nuclei and lost enough momentum became gravitationally trapped.
Typically a sufficient density of WIMPs has accumulated in the solar
core that equilibrium now exists between WIMP capture and
annihilation.  Then, given a WIMP mass and decay branching ratios, one
can unambiguously predict the signal for a neutrino telescope.  The
IceCube Collaboration continues to take data and has performed or is
performing searches for neutrino signals from WIMPs in the center of
the Earth~\cite{Achterberg:2006jf}, the solar
core~\cite{Aartsen:2012kia}, the galactic halo~\cite{Abbasi:2011eq}
and center~\cite{Abbasi:2012ws}, galaxy clusters, and
dSphs~\cite{IceCube:2011ae}.

The PINGU detector is proposed as a new in-fill array for IceCube.
PINGU will instrument an effective volume of several million metric
tons for neutrinos with energy $E_\nu \sim$ 5--15~GeV.  The final
geometry for the detector is still under study but will probably be
comprised of 20 to 40 new strings with 60--100 modules per string.
PINGU is focused on other physics goals but has sensitivity to
neutrinos produced by WIMP annihilations.  Together IceCube and PINGU
will be able to probe a WIMP mass region that is of considerable
interest, given intriguing results from other experiments that are
consistent with a WIMP mass of a few GeV.  In particular, WIMP
properties motivated by DAMA's annual modulation
signal~\cite{Savage:2008er} and isospin-violating
scenarios~\cite{Feng:2011vu} motivated by DAMA and CoGeNT signals will
be tested.  The predicted sensitivities of Hyper-Kamiokande and PINGU
are shown in \Figref{ctasens2}.  Depending on the WIMP mass, IceCube
or PINGU could detect a smoking-gun signal of dark matter (a
high-energy neutrino signal from the Sun) as well as place competitive
limits on the spin-dependent nuclear recoil cross section, compared
with planned G2 direct detection experiments.

\subsection{Astrophysical multiwavelength constraints}
 
Pair annihilation of WIMPs will result in a conspicuous non-thermal
population of energetic electrons and positrons from the decays of
charged pions produced by the hadronization of strongly interacting
final-state particles, as well as from the decays of gauge bosons,
Higgs bosons, and charged leptons. This non-thermal $e^\pm$ population
loses energy and produces secondary radiation through several
processes, covering a wide range of the electromagnetic spectrum from
the radio to the $\gamma$-ray band.  Several recent studies have made
it clear that searches in the radio~\cite{Storm:2012ty} and X-ray
frequencies~\cite{Jeltema:2011bd} have the potential to reach
sensitivities to the relevant WIMP parameter space that are comparable
to, and in some instances broader than and complementary to, the
sensitivities of $\gamma$-ray experiments. A detailed performance
comparison depends critically on assumptions about propagation, energy
losses, and the astrophysical environment where the secondary
radiation is emitted~\cite{Storm:2012ty}.  An accurate definition of
benchmarks is crucial for this field.

\subsection{Indirect detection conclusions}

The primary methods of indirect detection include $\gamma$-ray
measurements of the center of our own galactic halo, in nearby dwarf
galaxies and in clusters; observation of charged cosmic-ray
antimatter; searches for high energy neutrinos from annihilation in
the Sun; and astrophysical signatures, such as radio and hard X-ray
emission.  Just as the next generation of direct detection experiments
is approaching the natural range for WIMP scattering cross sections,
$\gamma$-ray experiments are reaching the sensitivity required to
exclude the natural range of model predictions for WIMP annihilation
cross sections.  Given the close relation of the $\gamma$-ray
production cross section to the total annihilation cross section, the
bulk of the likely parameter space for SUSY WIMPs (or, in fact, any
weakly interacting massive thermal relic) is focused in a relatively
narrow band of cross sections compared to the nuclear recoil
scattering cross section.  Comparing direct and indirect detection,
the dominant systematic uncertainty for direct detection from the
nuclear recoil cross section might be even larger than the
uncertainties in indirect searches from halo models.  Even the
conservative, cored halos give an observable signal for future
$\gamma$-ray experiments.

A future experiment like CTA, with U.S. enhancement, could reach
most of the parameter space in well-explored theoretical frameworks,
such as the pMSSM, through observations of the GC for all but the most
pessimistic assumptions about the halo profile or new astrophysical
backgrounds.  U.S. involvement in CTA is critical.  This involvement
would result in a doubling of the planned mid-sized telescope array
with substantial improvements in sensitivity and angular resolution
that are much more than an incremental improvement.  Perhaps of
equal importance, involvement of the U.S. HEP community would provide
not only the technical expertise but also the scientific impetus to
devote a large fraction of CTA's observing time to dark matter
targets.

New discoveries of nearby dwarf galaxies and new analysis techniques
are likely to result in Fermi observations cutting into the pMSSM
parameter space to energies up to tens of GeV.  Above a few hundred
GeV, VERITAS will come within an order of magnitude of the natural
annihilation cross section for a combined dwarf analysis.  IceCube
searches for high-energy neutrinos from the Sun will continue to
provide some of the most sensitive constraints on the spin-dependent
scattering cross section and offer the potential for a smoking-gun
discovery.  The proposed PINGU extension to IceCube would provide
sensitivity to WIMP masses favored by DAMA.

This is an exciting time for dark matter research, where direct
detection experiments may see a hint of a signal, the LHC may find new
physics, $\gamma$-ray measurements may identify the particle and
measure its distribution in galactic halos, and neutrino measurements
may provide a smoking-gun detection from the annihilation signal from
the Sun.  Without all of these avenues for research, the story would
be incomplete.

\section{Non-WIMP dark matter}
\label{sec:nonwimp}

In solving the mystery of dark matter, it is sensible to assume the
properties of the dark matter particle candidate and predict its
non-gravitational interactions before the particle can be identified.
It is common to predict the properties of dark matter based on either
compelling theoretical arguments or experimental and astrophysical
hints.  The relative significance of different arguments and hints
cannot easily be evaluated objectively, which makes it difficult to
rank dark matter candidates in importance.  Although a combination of
certain theoretical arguments, advanced direct search experimental
techniques, and connection with collider experiments makes WIMPs
attractive candidates, they are by no means the only appealing
possibility.

Furthermore, there is no reason to believe that dark matter is
comprised of only one type of particle. It is possible that the
structure of the dark sector is as complex as that of the visible
sector, or even more complex. A broad, extensive, and multifaceted
approach to solving the mystery of dark matter is more likely to yield
exciting discoveries than a focused pursuit of one dark matter
candidate. At the same time, the enormous breadth of possibilities
forces us to identify some directions as the most promising.
Theoretical plausibility and experimental feasibility are commonly
used in the community for this purpose.

\begin{figure}[tbp]
\begin{center}
  \includegraphics[width=0.73\hsize]{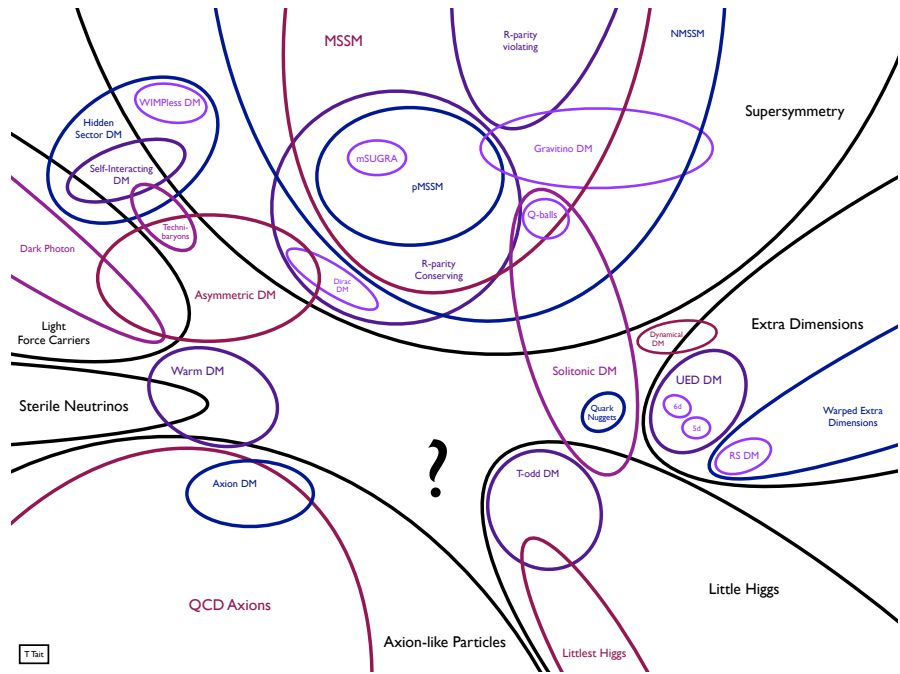} 
\caption{The landscape of dark matter candidates [from T.~Tait].
\label{fig:landscapevenn}}
\end{center}
\end{figure}

\begin{figure}[tbp]
\begin{center}
\includegraphics[width=0.73\hsize]{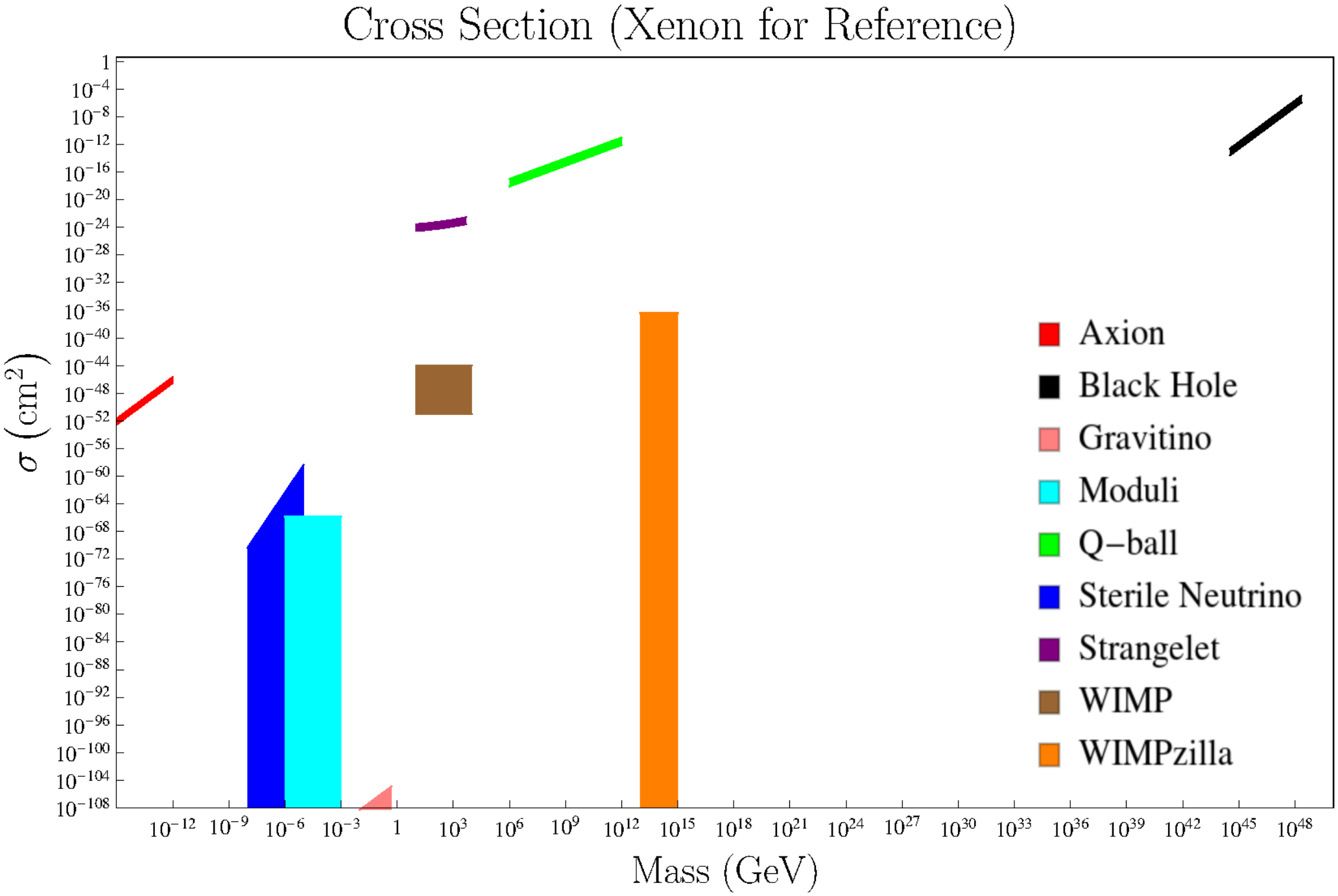}
\caption{The range of dark matter candidates' masses and interaction
  cross sections with a nucleus of Xe (for illustrative purposes)
  compiled by L.~Pearce. Dark matter candidates have an enormous range
  of possible masses and interaction cross sections.
\label{fig:landscape}}
\end{center}
\end{figure}

The landscape of non-WIMP candidates is sketched in
\Figref{landscapevenn}. There are a large number of possibilities.  To
be directly detected, the dark matter needs to interact in some way
with normal matter and radiation.  \Figref{landscape} shows the
incredible range of interaction strengths and masses of the different
candidates.  Generally, the lighter the candidate, the more feeble its
allowed interactions. Very heavy dark matter particles can have quite
strong interactions. This wide range of coupling strengths and masses
calls for an equally wide range of detector technologies. For the most
feeble interactions of the axion, the detection technology is based on
quantum-limited detection of radio photons. The rare interactions of
the more massive candidates, such as Q-balls, produce distinctive
signatures in large detectors.  Finally, a number of candidates cannot
be found in laboratory experiments in the foreseeable future, and so
X-ray and gamma-ray telescopes, as well as gravitational lensing
techniques, provide the best opportunities for their detection.

The following non-WIMP dark matter candidates were highlighted in the
course of Snowmass.  This list is not a complete list of possible
candidates, but it includes candidates for which there was substantial
interest and identifiable experimental and theoretical research paths,
and which were discussed in white paper contributions.

\begin{itemize}

\item {\bf Asymmetric dark matter} is motivated by the fact that the
  abundances of ordinary matter and dark matter in the Universe are
  comparable~\cite{Petraki:2013wwa,Zurek:2013wia}.  The two quantities
  can be similar and related if the two types of matter are produced
  simultaneously, or if they are generated from two separate but
  similar processes in the early Universe.  Since the abundance of
  ordinary matter is controlled by the baryon asymmetry of the
  Universe, a similar asymmetry may be responsible for the observed
  dark matter abundance.  In contrast with WIMPs, asymmetric dark
  matter candidates can have masses and interaction cross sections
  that are very different from the naive expectations motivated by the
  freezeout of a thermal relic.

\item {\bf Axions} arise from an elegant (and, arguably, the only
  viable) solution to the strong CP problem in the Standard
  Model~\cite{Peccei:1977hh,Peccei:1977ur,Weinberg:1977ma,Wilczek:1977pj}.
  The vacuum of quantum chromodynamics (QCD) is characterized by a
  parameter $\theta_{\rm QCD}$, which is independent of the phase
  $\theta_{\rm q}$ in the quark mass matrix.  It is a mystery why the
  sum of these two unrelated parameters, which controls CP violation
  in strong interactions, should vanish to an experimentally-measured
  precision of one part in ten billion.  Peccei--Quinn theory explains
  this mystery and predicts a new light degree of freedom, the axion,
  which can also be dark matter.  There are several promising
  experimental techniques to search for axions, and this program was
  viewed to have a strong discovery potential. A variant candidate is
  the axion-like-particle, a light particle with the axion's quantum
  numbers, but having mass and couplings not linked by the
  Peccei-Quinn theory.  Searches for such particles were discussed by
  both Cosmic Frontier and Intensity Frontier working groups.

\item {\bf Primordial black holes} can form in the early
  Universe~\cite{Hawking:1971ei,Khlopov:1985jw,Khlopov:2008qy}, for
  example, from the same dynamics that govern cosmological
  inflation~\cite{Frampton:2010sw}.  Although theoretical models
  invoke fine-tuning of parameters to explain the dark matter
  abundance, primordial black holes are a viable dark matter candidate
  that can be discovered using gravitational lensing
  observations~\cite{Cieplak:2012dp}.

\item {\bf Self-interacting dark matter} can explain some
  inconsistencies between the predictions of $N$-body simulations and
  observations~\cite{Spergel:1999mh,Rocha:2012jg,Peter:2012jh,%
    Vogelsberger:2012ku,Zavala:2012us}.

\item {\bf Sterile neutrinos} are motivated by the fact that active
  neutrinos are massive, which is typically explained by introducing
  right-handed, gauge-singlet
  fermions~\cite{Dodelson:1993je,Kusenko:2009up}.  If one or more of
  these fermions is relatively light, as occurs in a number of
  different
  models~\cite{Asaka:2005pn,Kusenko:2006rh,Kusenko:2010ik,Merle:2013gea},
  the dark matter can consist of sterile neutrinos.  In a part of the
  parameter space allowed for dark matter, the same sterile neutrinos
  can explain the observed pulsar velocities, since these sterile
  neutrinos would be emitted anisotropically from a cooling neutron
  star born in a supernova
  explosion~\cite{Kusenko:1997sp,Fuller:2003gy}.  The most promising
  detection strategy is based on the radiative decays of sterile
  neutrinos, which can produce a line detectable by X-ray
  telescopes~\cite{Abazajian:2001vt}.

\item {\bf Superheavy dark matter} must have a low number density,
  which requires the use of very large detectors.  Indirect detection
  of decay products can lead to discovery.  Several candidates fall in
  this category, including quark
  nuggets~\cite{Witten:1984rs,Zhitnitsky:2002qa} and
  WIMPzillas~\cite{Chung:1998ua}.

\item {\bf Supersymmetric non-WIMP candidates}, including SUSY
  Q-balls~\cite{Kusenko:1997zq,Kusenko:1997si} and products of their
  decays, as well as superWIMP dark matter~\cite{Feng:2003xh} are
  well-motivated dark matter candidates in theories with
  supersymmetry.  The search for these candidates is possible using
  direct and indirect techniques targeting specific properties of
  these candidates.

\end{itemize}

Among all these candidates, the axion deserves special mention.  For a
dark matter candidate, the QCD axion has the rare advantage of a
fairly well-bounded parameter space. Although there are ways to evade
the bounds, the axion-photon couplings $g_{a\gamma\gamma}$ over the
range of benchmark models extend over an order of magnitude. The upper
end of the QCD axion mass range is set at a few meV by the limit from
SN1987A, and the lower end, limited by the requirement that axions not
overclose the Universe, is set at around a $\mu$eV.  A particularly
promising approach to detect the QCD axion is via the RF-cavity
technique~\cite{Asztalos:2006kz}.  Although the expected conversion
into RF power within the cavity is extraordinarily weak, experiments
will shortly start taking data for a definitive search, which will
either find the QCD dark matter axion with high confidence, or exclude
it at high confidence. These experiments will sensitively explore the
first two decades of allowed QCD axion mass, where the dark matter QCD
axion is expected to be found. These searches have a large discovery
potential over the next decade.  There are axion and
axion-like-particle alternatives to the QCD axion, and this opens a
vast and largely unexplored search space. Much of this space,
including the third decade of allowed QCD axion mass, could be
explored by large next-generation detectors. For instance, the
proposed solar-axion experiment IAXO would explore this space,
including where there are astrophysical hints of new physics, yielding
a good discovery potential.  The axion might be found anywhere within
the allowed parameter space, and there are arguments for both
higher-mass axions (which would then be of the non-Peccei-Quinn type)
or lower-mass axions (which could be QCD axions whose primordial
abundance is explained by anthropic selection).

It may also be that the dark matter consists of several different
particles.  The range of possibilities is enormous. Aside from axions,
proposals for special-purpose non-WIMP dark matter detectors and
programs are less well-developed.  However, existing experiments and
astronomical instruments offer a number of serendipitous
opportunities.

Once the axion or other dark matter particle is identified, it will
mark a new beginning.  For instance, one virtue of the RF-cavity
experiments is that they measure the total energy of the axion ---
mass plus kinetic energy --- and there may be fine structure to the
signal due to the flows of dark matter in the halo.  This contains a
wealth of information about the history of the formation of our Milky
Way galaxy and will mark the beginning of a new field of
astronomy~\cite{Sikivie:1996nn}.  On the other hand, if dark matter is
made up of sterile neutrinos, the narrow spectral line from their
decay could provide information about the redshift, allowing us to map
out dark matter in the Universe and to use the redshift information
for measuring the cosmological expansion~\cite{Kusenko:2009up}.  Much
the same can be said about discovering any of the dark matter
candidates: The identification of dark matter will be a revolutionary
discovery that will open the door to a new chapter in our
understanding of nature.

\section{Dark matter complementarity} 
\label{sec:complementarity}

All current evidence for dark matter is derived solely through its
gravitational interactions.  The presence of dark matter has been
quantified on length scales that range from the solar neighborhood to
the horizon of the Universe. The history extends back to Jan Oort, who
proposed using stars above the plane of the Milky Way as a way to
minimize the influence of the stellar disk~\cite{1932BAN.....6..249O},
and Fritz Zwicky who postulated dark matter in 1937 from the large
measured velocity dispersion of galaxies in the Coma
cluster~\cite{1937ApJ....86..217Z}. Subsequent progress in the 1970's
in measuring the rotation curves of nearby galaxies in both optical
and radio frequencies~\cite{1979A&A....79..281B, 1980ApJ...238..471R}
really brought the idea of dark matter home to many astronomers and
physicists. The support for dark matter in the form of cold relic
particles came about through revolutions in cosmology including galaxy
correlations~\cite{Tegmark:2003ud,Hawkins:2002sg}, the
CMB~\cite{Komatsu:2010fb,Ade:2013lta}, and measurements of hot gas in
clusters~\cite{Allen:2002eu}. Visceral evidence for dark matter is
provided by strong and weak lensing
measurements~\cite{Refregier:2003ct,Tyson:1998vp}, including the
famous Bullet Cluster~\cite{Clowe:2006eq}.  Together, these data
provide overwhelming evidence that the energy in dark matter is
roughly a quarter of the total energy in the visible Universe and
about five times the energy in normal matter.

To understand dark matter, we need to answer some basic questions: (1)
How many particle species make up the dark matter? (2) What are their
masses and spins? (3) How do they couple to the Standard Model and
other new (dark sector) particles? Our primary conclusion is that the
broad range of possibilities for dark matter (more than one of which
may be correct) argues strongly for multiple search strategies,
including direct searches, indirect searches, collider searches, and
astrophysical probes, to answer these basic questions.

The Dark Matter Complementarity working group highlighted the
essential synergies of these different approaches. These approaches,
as well as the complementarity of experiments within each approach,
are discussed in greater depth above in Secs.~\ref{sec:direct},
\ref{sec:indirect}, and \ref{sec:nonwimp}, and in the reports of the
Cosmic Frontier dark matter working groups and the Energy Frontier new
physics working group.

\subsection{Dark matter candidates and the need for a 
multi-pronged search strategy}
\label{sec:candidates}

Several classes of particles are strong dark matter candidates. The
most familiar dark matter candidates are WIMPs, discussed in
\Secsref{direct}{indirect}, which are produced in the hot early
Universe and then annihilate in pairs.  Those that survive to the
present are known as ``thermal
relics''~\cite{Zeldovich:1965,Chiu:1966kg,Steigman:1979kw,Scherrer:1985zt}.
Such particles are generically predicted in models of physics beyond
the Standard Model, including models with
supersymmetry~\cite{Goldberg:1983nd,Ellis:1983ew} or extra spatial
dimensions~\cite{Servant:2002aq,Cheng:2002ej}.  If these particles
interact through the weak interactions of the Standard Model, the
resulting thermal relic density is close to the observed dark matter
density.  This coincidence, known as the ``WIMP miracle,'' provides
strong motivation for dark matter with masses up to about a TeV.

As discussed in \Secref{nonwimp}, however, there are also many
alternative scenarios. For example, in the case of asymmetric dark
matter~\cite{Nussinov:1985xr,Gelmini:1986zz,Barr:1991qn,%
  Kaplan:1991ah,Kaplan:2009ag}, there is a slight excess of dark
particles over dark antiparticles in the early Universe.  These
annihilate until only the slight excess of dark particles remains.  In
many models, the dark matter asymmetry is related to the normal
matter--antimatter asymmetry, and one expects the number of dark
matter particles to be similar to the number of protons, implying a
dark matter particle mass of $\sim 10~\gev$, assuming this particle is
the dominant component of dark matter.  Asymmetric or thermal relic
dark matter may be in a so-called hidden sector, which has its own set
of matter particles and forces, through which the dark matter
interacts with other currently unknown particles.

\begin{figure}[tb]
\centerline{\includegraphics[width=0.93\hsize]{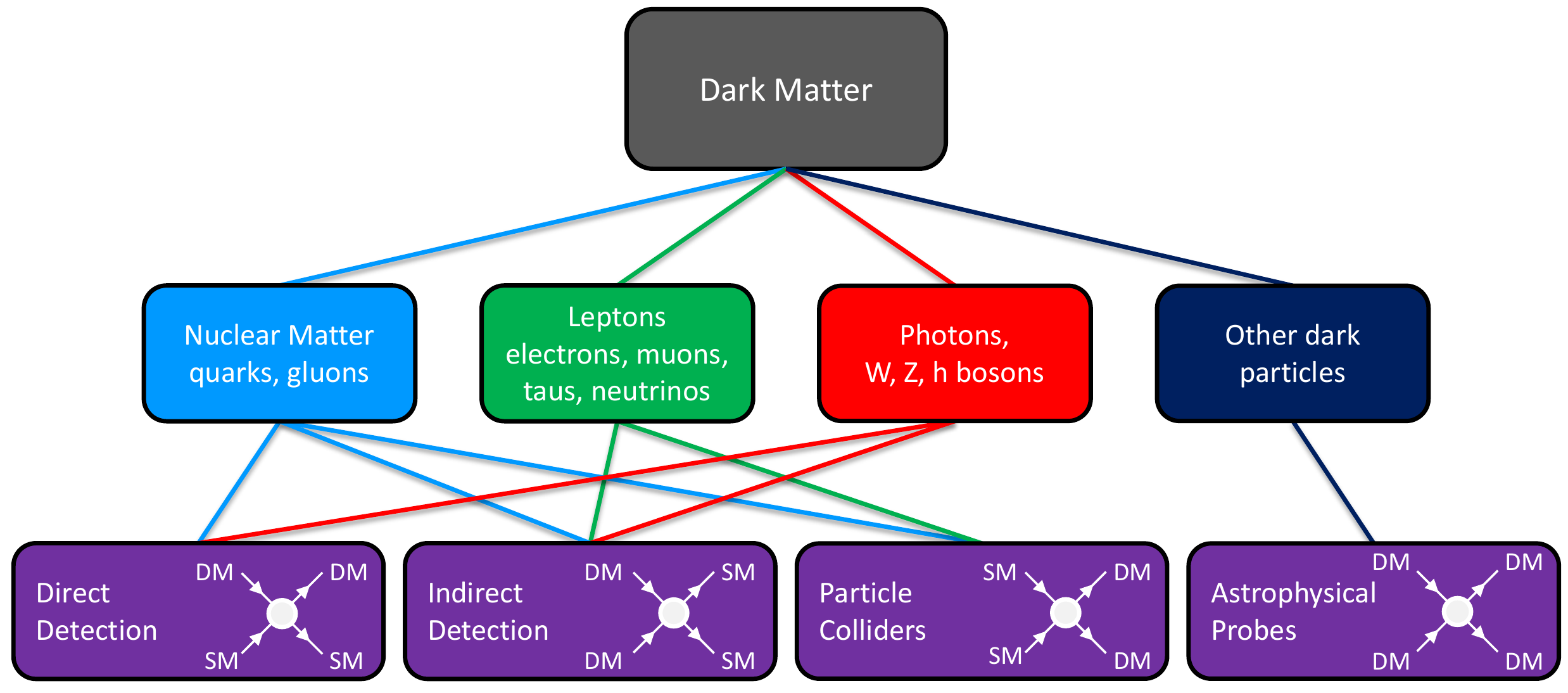}}
\vspace*{.2in}
\caption{Dark matter may have non-gravitational interactions with one
  or more of four categories of particles: nuclear matter, leptons,
  photons and other bosons, and other dark particles.  These
  interactions may then be probed by four complementary approaches:
  direct detection, indirect detection, particle colliders, and
  astrophysical probes. The lines connect the experimental approaches
  with the categories of particles that they most stringently probe.
  The diagrams give example reactions of dark matter (DM) with
  Standard Model particles (SM) for each experimental approach. From
  Ref.~\cite{Bauer:2013ihz}.
\label{fig:interactions}}
\end{figure}

The non-gravitational interactions of the above dark matter candidates
may be with any of the known particles or, as noted above for hidden
sector dark matter, with other currently unknown particles.  A
complete research program in dark matter therefore requires a diverse
set of experiments that together probe all possible types of
couplings, as shown in \Figref{interactions}.  At a qualitative level,
the complementarity may be illustrated by the following observations
that follow from basic features of each approach.

\begin{itemize}

\item {\bf Direct Detection.}  Prime examples are the detection of
  WIMPs through scattering off nuclei and the detection of axions in
  RF cavities. The methods relying on scattering off nucleons are
  relatively insensitive to dark matter that couples to leptons only,
  or to WIMP-like dark matter with mass $\sim 1~\gev$ or below, while
  they are powerful for dark matter that couples to charged particles
  or gluons.

\item {\bf Indirect Detection.}  When pairs of dark matter particles
  annihilate, they produce high-energy particles in the cosmic rays.
  For example, antimatter from local annihilation events can be found
  by AMS-02, neutrinos from annihilations in the Sun can be detected
  at IceCube, and photons from annihilations at the GC or in other
  galaxies can be seen by $\gamma$-ray telescopes.  Alternatively,
  dark matter may be metastable, and its decay may produce the same
  high-energy particles. These indirect detection experiments
  (together) are sensitive to the interactions with all Standard Model
  particles and probe the annihilation process suggested by the WIMP
  miracle. Experimental sensitivities are expected to improve greatly
  on several fronts in the coming decade but some modes require good
  understanding of astrophysical backgrounds. Further, the signals are
  typically subject to uncertainties in the spatial distribution of
  dark matter (which is often not directly constrained) and may be
  absent altogether whenever the dark matter annihilation is
  insignificant now, \eg, in the case of asymmetric dark matter or
  $P$-wave suppressed annihilation.

\item {\bf Particle Colliders.}  Particle colliders, such as the
    Large Hadron Collider (LHC) and proposed future lepton colliders,
    produce dark matter particles that escape the detector, but are
    discovered as an excess of events with missing energy or momentum.
    LHC experiments are sensitive to the broad range of masses favored
    for WIMPs (especially if they couple to quarks and/or gluons), but
    are relatively insensitive to dark matter that interacts only with
    leptons. Collider experiments are also unable to distinguish
    missing momentum signals produced by a particle with lifetime
    $\sim 100$ ns from one with lifetime above $10^{17}~{\rm s}$, as
    required for dark matter.

\item {\bf Astrophysical Probes.}  The particle properties of dark
  matter are constrained through its impact on astrophysical
  observables.  Dark matter distributions and substructure in galaxies
  are unique probes of the ``warmth'' of dark matter and hidden dark
  matter properties, such as its self-interaction strength, and they
  measure the effects of dark matter properties on structure formation
  in the Universe.  Examples include the self-interaction of dark
  matter particles affecting central dark matter densities in galaxies
  (inferred from rotation velocity or velocity dispersion measures),
  the mass of the dark matter particle affecting dark matter
  substructure in galaxies (inferred from strong lensing data), and
  the annihilation of dark matter in the early Universe affecting CMB
  fluctuations.  Astrophysical probes are typically unable to
  distinguish various forms of CDM from one another or make other
  precision measurements of the particle properties of dark matter.
\end{itemize}

\subsection{Discovery complementarity}

\begin{figure}[tbp]  
\includegraphics[height=0.45\columnwidth]{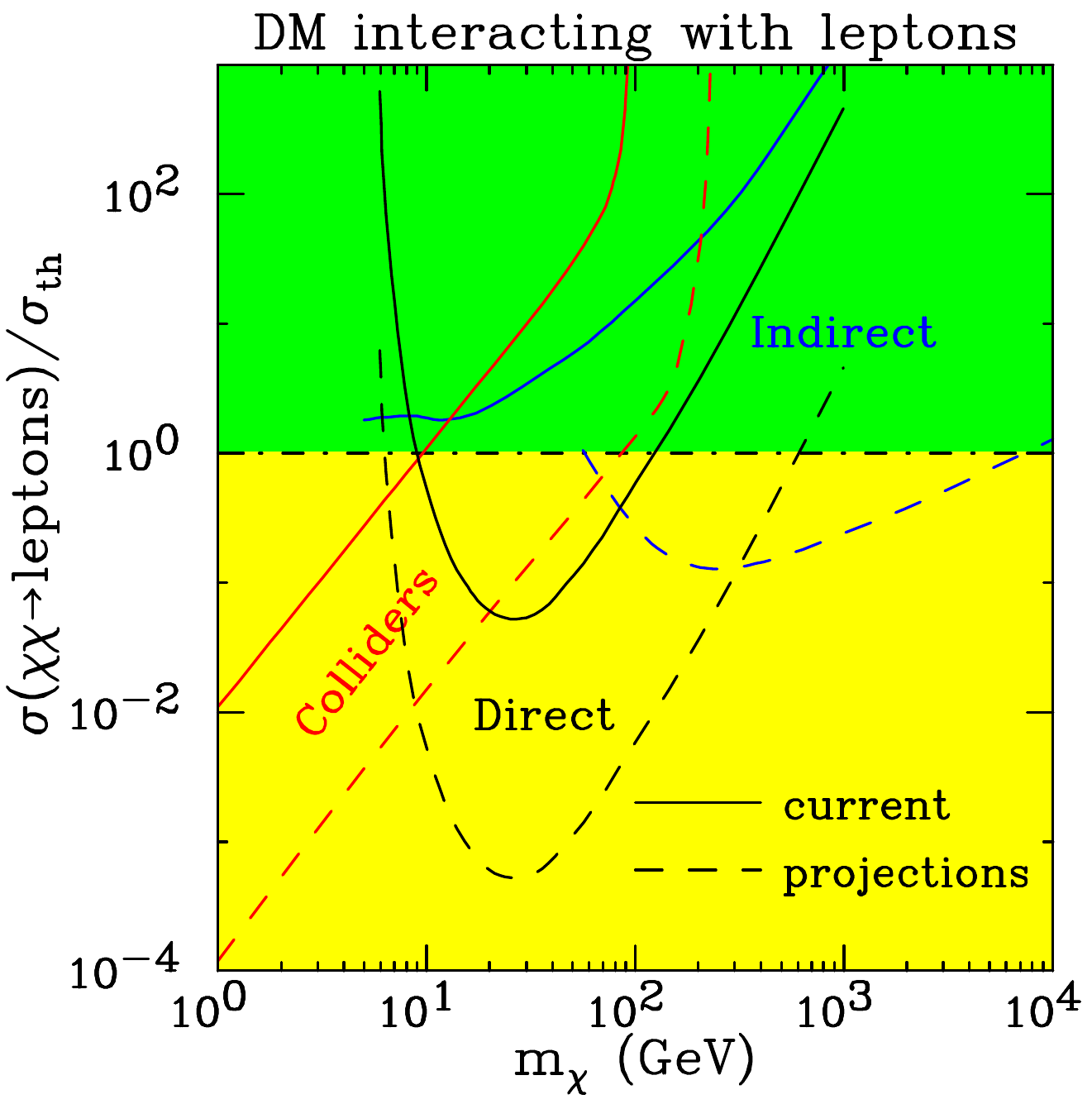}
\includegraphics[height=0.45\columnwidth]{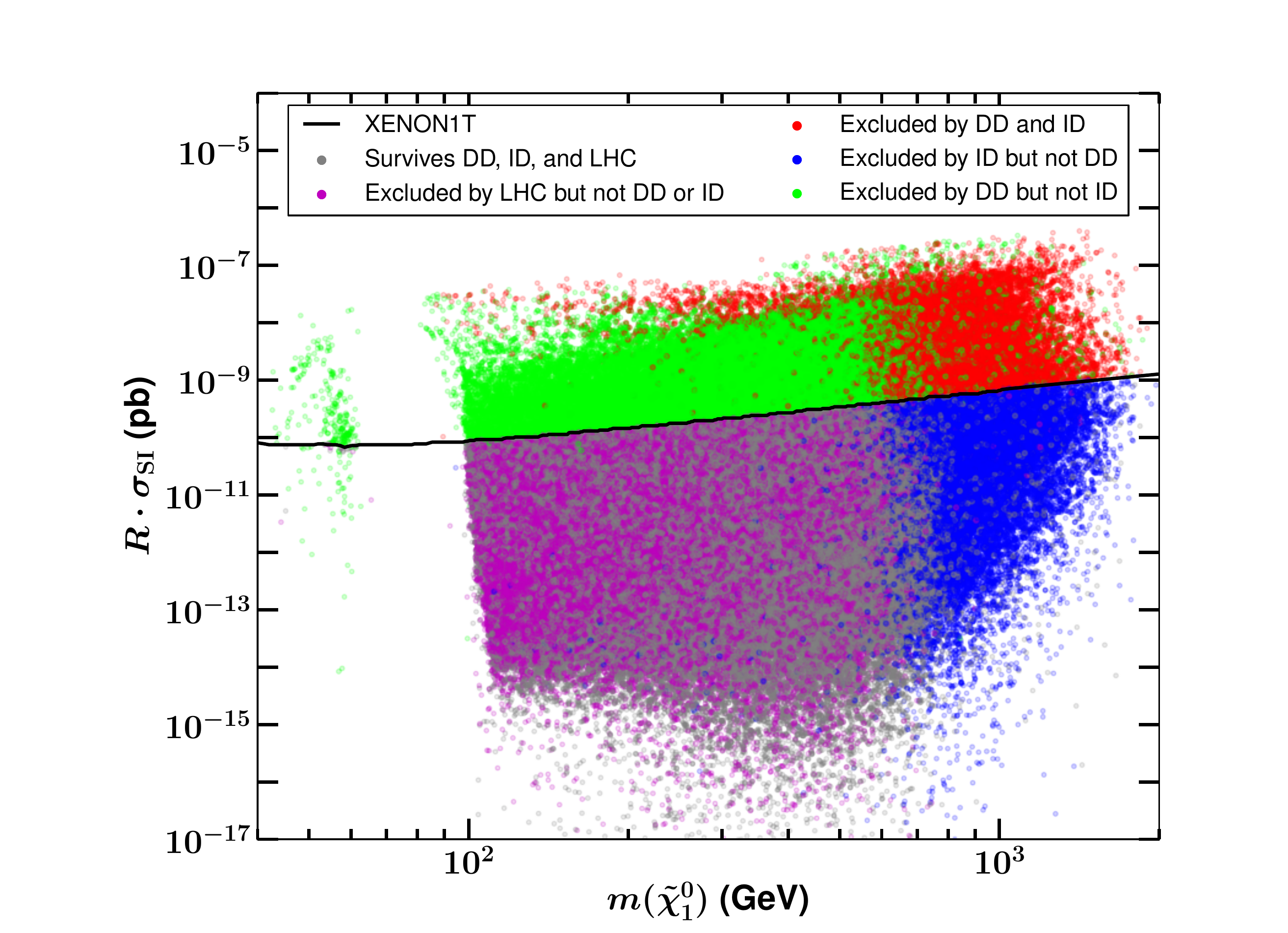} \\
\hspace*{1.3in}
\includegraphics[height=0.55\columnwidth]{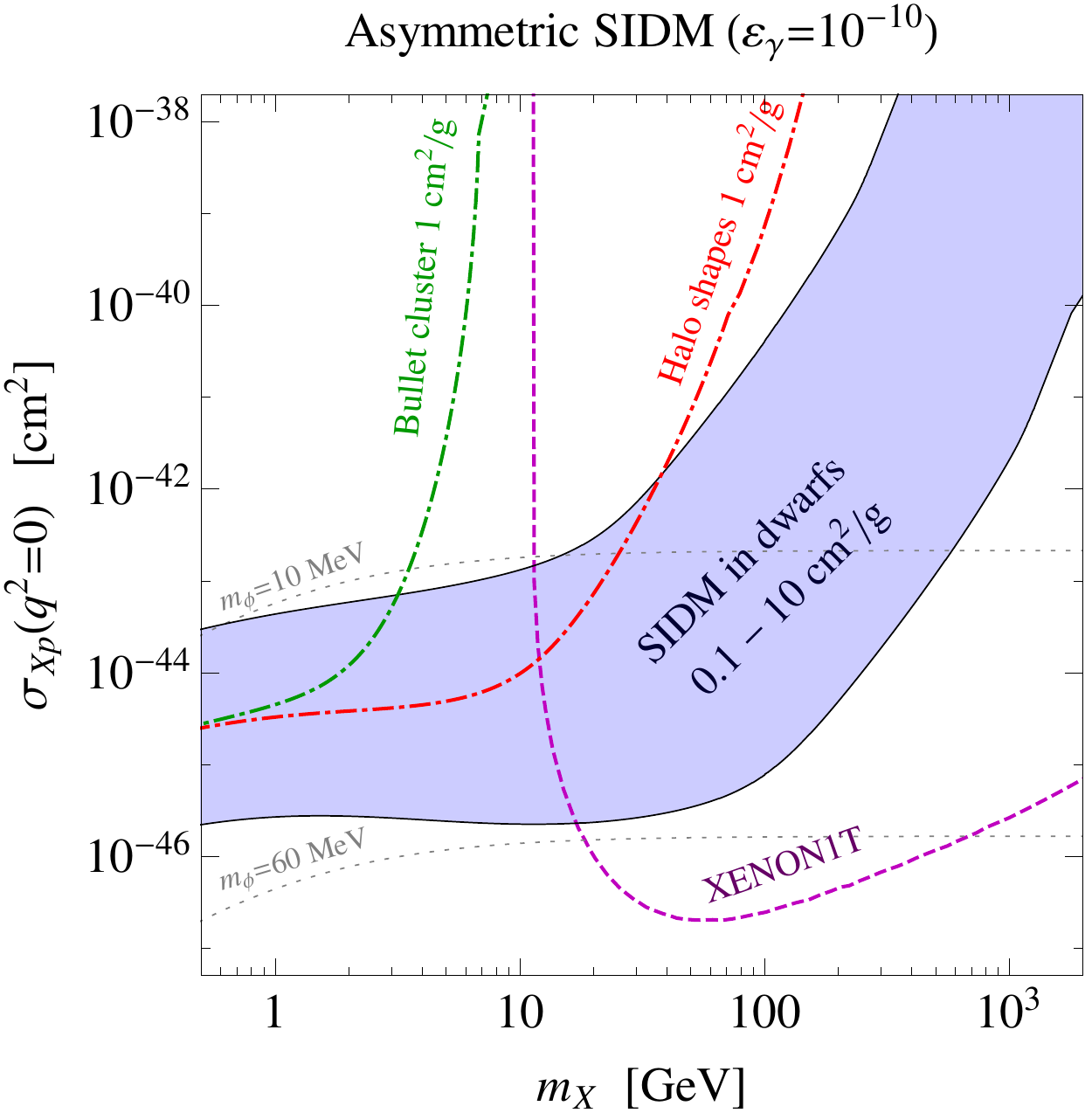}
\caption{Top left: Dark matter discovery prospects in the mass vs
  annihilation cross section plane in the effective operator framework
  for dark matter coupling to leptons~\cite{Bauer:2013ihz}. Top right:
  A model-independent scan of the full parameter space in the minimal
  supersymmetric model (MSSM), presented in the mass vs. dark
  matter-nucleon spin-independent cross section
  plane~\cite{Cahill-Rowley:2013dpa}.  Bottom: Prospects for direct
  detection of asymmetric dark matter that couples to quarks via gauge
  kinetic mixing with $\epsilon=10^{-10}$, showing complementarity
  with astrophysics in the mass vs.~dark matter-nucleon
  spin-independent cross section plane~\cite{Kaplinghat:2013kqa}.}
\label{fig:prospects}
\end{figure}

We illustrate the qualitative features outlined above in three
ways. First, we consider a fairly model-independent setting by
considering dark matter that interacts with Standard Model particles
through four-particle contact operators. These interactions are
expected to work well to describe theories in which the exchanged
particle mass is considerably larger than the momentum transfer of the
physical process of interest.  To illustrate aspects of the
complementarity of the different searches, we assume the dark matter
is a spin-1/2 particle that couples in a generation-independent way to
the Standard Model leptons (effective operators involving quarks or
gluons were also considered in Ref.~\cite{Bauer:2013ihz}).  The
relevant piece of the Lagrangian is $M_\ell^{-2}\bar{\chi} \gamma^\mu
\chi \sum_\ell \bar{\ell} \gamma_\mu \ell$, where $M_{\ell}$ is the
scale of the new physics. The complementarity of different
experimental approaches in probing low masses ($\le 10$ GeV), moderate
masses, and high masses ($> 300$ GeV) is evident in the top left panel
of \Figref{prospects}, which shows dark matter discovery prospects in
the mass vs.~annihilation cross section plane for current and future
direct detection experiments~\cite{DMtools}, indirect detection
experiments~\cite{Ackermann:2011wa,IndirectCTA}, and particle
colliders~\cite{Fox:2011fx,Chae:2012bq} for dark matter coupling to
leptons. A discovery in the yellow region of this plot (below the
dot-dashed line) would indicate that the dark matter (if assumed to be
a thermal relic) has other significant annihilation channels that are
waiting to be discovered.

The interactions with quarks and gluons are here illustrated through a
concrete example of a complete supersymmetric model. The dark matter
candidate is the lightest neutralino $\tilde\chi^0_1$, which is its
own antiparticle. Even within the general framework of supersymmetry,
there are many different model scenarios, distinguished by a number of
input parameters ($\sim 20$). A model-independent approach to
supersymmetry is to scan over all those input parameters and consider
models that pass all existing experimental constraints and have a dark
matter candidate that could explain at least a portion of the observed
dark matter density~\cite{CahillRowley:2012cb}.  Results from such
model-independent scans with over 200,000 points are shown in the
top right panel of \Figref{prospects}, where each dot represents one
particular supersymmetric model and its color indicates whether the
model is observable in future direct searches (green), indirect
searches (blue), or in both (red).  The models that escape both direct
and indirect searches, but are discoverable with current LHC data, are
shown in magenta.  The top right panel of \Figref{prospects} demonstrates
that these three different dark matter probes combine to discover a
large fraction of the supersymmetry models in this scan.

The bottom panel of \Figref{prospects} is for an asymmetric dark
matter candidate that is charged under a hidden sector broken U(1)
that has kinetic mixing with the Standard Model photon.  The broken
U(1)'s fine-structure constant is taken to be $10^{-2}$ for
illustration and the hidden sector gauge boson's mass is in the 1--100
MeV range.  Bounds from supernova cooling and beam dump experiments
restrict the value of the kinetic mixing parameter to be below about
$\epsilon = 10^{-10}$.  For this mixing parameter, the gauge boson
also safely decays before big bang nucleosynthesis.  The dark matter
self-interaction cross section in the blue shaded region is large
enough (0.2 to 20 barns/GeV) to affect dwarf galaxies on observable
scales. The effect is to lower the densities (see, \eg,
Ref.~\cite{2012MNRAS.422.1203B}) and create constant density cores
(see, \eg, Refs.~\cite{2008ApJ...676..920K,2011AJ....141..193O}) in a
manner that is preferred by observations.  Also shown are bounds from
Bullet Cluster observations and requirements that self-interactions
not make galactic halos too round; consistency with observations
excludes the parameter space above the indicated contours.  Direct
detection is sensitive to a large region of parameter space that is
preferred by astrophysical observations as shown by the contour for
XENON1T sensitivity. We see that astrophysical observables probe low
masses, while direct detection experiments probe mid-range and high
masses, illustrating the complementarity of these two approaches in
this asymmetric dark matter framework.  The same model also allows for
symmetric dark matter that freezes out with the right relic density
(like WIMPs), where the qualitative complementarity features are the
same but with further implications for indirect searches.

\subsection{Post-discovery complementarity} 

As important as a broad program of complementary searches is to
establishing a compelling signal for dark matter, for several reasons
it becomes even more important after a signal has been reported.

First, as is well known, many tentative dark matter signals have
already been reported.  The potential identification of a quarter of
the Universe will require extraordinary proof in the form of
verification by other experiments. Second, each search strategy has
its limitations, as noted earlier in \Secref{candidates}.  Last, the
discovery of dark matter will usher in a rich and decades-long program
of dark matter studies, requiring a multi-pronged approach.

Consider the following scenario: Direct search experiments see
evidence for a 100 GeV WIMP, and missing energy searches at an
upgraded LHC also see a consistent signal. However, further LHC and
ILC studies constrain the neutralino's predicted thermal relic density
to be half of $\Omega_{\rm DM}$, implying that it is not a thermal
  relic, or that it makes up only half of the dark matter.  The puzzle
  is resolved when axion detectors discover a signal that is
  consistent with axions making up the rest of the dark matter.
  Progress in astrophysical theory, simulations, and observations
  eventually leads to a consistent picture with dark matter composed
  entirely of CDM and extends our understanding of the Universe back
  to 1 ns after the big bang.  Direct and indirect detection rates are
  then used to constrain the local dark matter density, halo profiles,
  and substructure, establishing the new fields of neutralino and
  axion astronomy. We see from the above example that even for
  well-studied dark matter candidates, information from multiple
  approaches is required to fully understand the dark matter. This is
  even more true if the dark matter sector is richer than that
  imagined above.

A balanced program with components in each of the four approaches is
required to cover the many well-motivated dark matter possibilities,
and their interplay will likely be essential to realize the full
potential of upcoming discoveries.

\section{Dark energy and CMB} 

Maps of the Universe when it was 400,000 years old, from observations
of the CMB and over the last ten billion years from galaxy surveys,
point to a compelling cosmological model. This model requires a very
early epoch of accelerated expansion, known as inflation, during which
the seeds of structure were planted via quantum mechanical
fluctuations. These seeds began to grow via gravitational instability
during the epoch in which dark matter dominated the energy density of
the Universe, transforming small perturbations laid down during
inflation into nonlinear structures such as million-light-year-sized
clusters, galaxies, stars, planets, and people. Over the past few
billion years, we have entered a new phase, during which the expansion
of the Universe is accelerating, presumably driven by yet another
substance, dark energy.

Cosmologists have historically turned to fundamental physics to
understand the early Universe, successfully explaining phenomena as
diverse as the formation of the light elements, the process of
electron-positron annihilation, and the production of cosmic
neutrinos. However, the Standard Model of particle physics has no
obvious candidates for inflation, dark matter, and dark energy. The
amplitude of the perturbations suggests that the natural scale for
inflation is at ultra-high energies\footnote{Roughly, the measured
  amplitude of the density perturbations $\delta\rho/\rho \simeq
  10^{-5} \sim (E_{\rm inf}/m_{\rm Planck})^2/\sqrt{\epsilon}$, where
  $\epsilon\simeq 0.01$ is a small parameter in slow roll inflation.},
so understanding the physics driving inflation could lead to
information about the ultraviolet completions of our current
theories. There are arguments that naturally link the dominant dark
matter component to new physics hovering above the electroweak scale;
the powerful experiments aiming to find this component have been
discussed above in Secs.~\ref{sec:direct}, \ref{sec:indirect},
\ref{sec:nonwimp}, and \ref{sec:complementarity}. Apart from the
dominant component, neutrino oscillation experiments already inform us
that neutrinos constitute a non-negligible fraction of the dark
matter.  One of the key points of this chapter is that experiments
usually associated with dark energy and inflation are ideally suited
to pin down the sum of the masses of the neutrinos and the cosmic
existence of any additional (sterile) species.

The situation with dark energy is more complex.  A cosmological
constant ($\Lambda$) has effective pressure and energy density related
by $p = -\rho$ (equation of state $w=-1$), consistent with preliminary
measurements, but in supersymmetric theories, for example, the most
natural scale for $\Lambda$ is at least 100 GeV. A cosmological
constant with this value would produce a Universe accelerating so
rapidly that the tips of our noses would be expanding away from our
faces at a tenth the speed of light. If $\Lambda$ is responsible for
the current epoch of acceleration, its value is many orders of
magnitude smaller than this, but curiously just large enough that it
began dominating the energy density of the Universe only recently. The
mechanism driving the current accelerated expansion of the Universe
remains a profound mystery.

The quest to understand dark energy, dark matter, and inflation is,
then, driven by a fundamental tension between the extraordinary
success of the model that explains our Universe and the failure of the
Standard Model of particle physics to provide suitable candidates for
the dark sector that is so essential to our current view of the
Universe. Experiments on the Cosmic Frontier have demonstrated that
the Standard Model is incomplete; the next generation of experiments
can provide the clues that will help identify the new physics
required.  

\subsection{Dark energy}

Physicists have proposed a number of different mechanisms that could
be responsible for the accelerated expansion of the Universe.  None is
compelling, but some of them have been predictive enough to fail,
while others have led to a deeper understanding of field theory and
gravity. Apart from the cosmological constant solution itself, all are
predicated on the assumption that there is some (unknown) mechanism
that sets $\Lambda$ to zero.

One possibility is that there is a previously undiscovered substance
that contributes to the energy density of the Universe in such a way
that the expansion accelerates. In the context of GR, acceleration
(the positive second derivative of the scale factor $a$) is governed
by Einstein's equations, which reduce to
\begin{equation}
\frac{\ddot a}{a} = - \frac{4\pi G}{3}\,\left[ \rho + 3P\right] \ ,
\end{equation}
where $G$ is Newton's constant, $\rho$ the energy density, and $P$ the
pressure.  The substance that drives acceleration, therefore, must
have negative pressure, or equation of state $w\equiv P/\rho <
-1/3$. A nearly homogeneous scalar field whose potential energy
dominates over its kinetic energy satisfies this requirement, leading
to {\em quintessence} models with potentials designed to fit the
data. In almost all viable models, however, the mass of the field must
be less than the Hubble scale today, of order $10^{-33}$ eV. Embedding
such a field in some extension of the Standard Model therefore is
challenging, as one would expect the scalar mass to get loop
corrections many orders of magnitude larger than this. This mass
stability problem is an indication of just how hard the problem is:
The mass that is protected is some 44 orders of magnitude smaller than
the Higgs mass that lies at the center of the electroweak hierarchy
problem. It seems clear that the new physics is an infrared
phenomenon, as opposed to all prior hints of new physics, which have
entered from the ultraviolet.

With no real guidance from theory, quintessence models are
nevertheless appealing because they open up the parameter space: Most
models have $w\ne -1$, and many have evolving equations of state so
that $dw/da$ is also non-zero. This class of models therefore offers a
clear, and arguably more appealing, alternative to the cosmological
constant that would be favored if surveys find a deviation from
$w=-1$.  Although the quintessence field does not clump on scales
smaller than its (very large) Compton wavelength, it should clump on
the largest scales. This is yet another difference from the
cosmological constant model. Finally, some models allow for episodic
dark energy domination, so that the present accelerating era is not
particularly special. Indeed, an early epoch of inflation is one such
epoch, but there may have been others, for example, during phase
transitions. Measuring the effects of dark energy in a series of
redshift bins is therefore necessary to distinguish among the many
possibilities.\footnote{Redshift $z\equiv a^{-1} - 1$; high redshift
  corresponds to early cosmic time.}  This is an area where the CMB,
whose lensing maps are sensitive to structure from redshift $z=10^3$
until today, can be profitably combined with galaxy surveys, whose
lensing maps probe structure at a sequence of lower redshifts.

Quintessence models explicitly introduce an extra degree of freedom in
the form of a new scalar field. An alternative is to modify Einstein's
theory of gravity so that it produces acceleration even in the absence
of dark energy. Early attempts~\cite{Carroll:2003wy} to modify gravity
in this way (the ``left-hand side of Einstein's equations'') also
implicitly introduced an extra degree of freedom~\cite{Chiba:2003ir},
but in a different way than do quintessence models.  Technically,
these models differ from quintessence in that the coefficient of the
Ricci scalar $R$ in the Jordan frame in the action depends on the new
field.\footnote{Theories can be written either in the Jordan frame, in
  which matter couples to the metric only through
  $\sqrt{-g}\mathcal{L}_m$, or in the Einstein frame, in which the
  gravitational part of the action remains Einstein-Hilbert and there
  are additional couplings of the new degrees of freedom to matter.}
Different scalar-tensor models can therefore produce the variety of
predictions found in quintessence models, but they extend the
possibilities in a new direction: The non-canonical coupling to the
Ricci scalar propagates to the equations that govern the evolution of
perturbations. This leads to the general conclusion that modified
gravity models typically predict that structure grows at a different
rate (\eg, Ref.~\cite{Hu:2007pj}) than in models based on
GR. Differentiating between modified gravity and GR-based dark energy
therefore will require measurements of cosmological distances to pin
down the background expansion, and then measurements of the growth of
structure to distinguish between them.

Although there have been many proposals for how to modify gravity, the
most interesting recent development traces back to an idea first
proposed by Fierz and Pauli~\cite{Fierz:1939ix}, that the graviton has
non-zero mass. Qualitatively a massive graviton seems like an
appropriate way to decrease the strength of the gravitational force on
very large scales and hence to explain the acceleration of the
Universe. In practice, the theory runs into two problems, both related
to the fact that a massive spin-2 particle carries degrees of freedom
beyond those of the massless graviton. These extra degrees of freedom
typically lead to modifications of GR in the solar system,
modifications that are excluded by the tight limits on post-Newtonian
parameters. The second challenge for massive graviton models is to
avoid Boulware-Deser ghosts~\cite{Boulware:1973my}, the instability of
one of the extra degrees of freedom.

To satisfy the solar system constraints, the extra degrees of freedom
need to be screened, \ie, heavily suppressed by limiting their range
of interaction or effective coupling to matter in environments like
the solar system. Any successful modified gravity model needs a
screening mechanism. One possibility is Vainshtein
screening~\cite{Vainshtein:1972sx}, which arises due to
non-linearities in the Fierz-Pauli potential. In general, these
non-linear terms do not avoid the ghost problem, but
recently~\cite{Nicolis:2008in} the set of terms in the potential that
are safe from ghosts has been identified. These {\em Galileon} models
offer a potentially attractive way of addressing the acceleration of
the Universe within a consistent framework. Beyond this theoretical
breakthrough, the Vainshtein screening intrinsic to this model (and
other proposed mechanisms) opens up yet another axis of tests: How and
where do modified gravity theories transition to normal Newtonian
gravity in the solar system? The full suite of methods to test
modified gravity models is still under development.

The modern view of the formation of structure in the Universe has been
confronted with a growing array of precise tests, including both the
temperature and polarization anisotropy spectrum of the CMB; light
curves of distant supernovae; abundances of galaxy clusters;
clustering of galaxies, quasars, and Lyman alpha systems;
gravitational lensing; and cross-correlations between different pairs
of these observations. The basic framework, with inflation, dark
matter, and dark energy at its core, has been confirmed repeatedly
over the past decade. We need to keep pushing: Either the basic
picture will break or the agreement will become even more
remarkable. Thinking in parameter space, the next decade will enable
us to reduce the uncertainty in the equation of state $w$ by a factor
ten. Historically in physics, precision measurements of this sort have
been pursued for these ends: Will the simple theory hold up or will it
need to be replaced by something more profound? In the case of the
cosmic acceleration, where there is no appealing fiducial model, the
push for greater precision takes on an even greater importance.

Since scientists in the United States discovered evidence for cosmic
acceleration over a decade ago, the U.S. has been the leader in the
field of dark energy studies.  The 2006 Dark Energy Task Force
report~\cite{Albrecht:2006um} provided a systematic discussion of
experimental approaches to dark energy and identified a sequence of
``Stage III'' and ``Stage IV'' dark energy experiments to build on
those then in progress. Stage III surveys were designed to address
systematic uncertainties, with the goal of statistics-limited
constraints from four independent probes. Each of these probes will be
developed over the ensuing decade to the point at which Stage IV
surveys will enable sub-percent level consistency checks for all
probes, some of which will have reached their cosmic variance limit,
\ie, the limit imposed by the fact that we have only a single
observable Universe to probe.  This staged categorization was
reiterated in the more recent Community Dark Energy Task Force
Report~\cite{rocky3}, which particularly emphasized the importance of
complementing planned imaging experiments with spectroscopic
experiments.  \Figref{Facilities} illustrates the timelines for
several of the major dark energy experiments in which U.S. scientists
are playing an important or leading role. Further details provided in
a separate document~\cite{Weinberg:2013raj}.

\Sfig{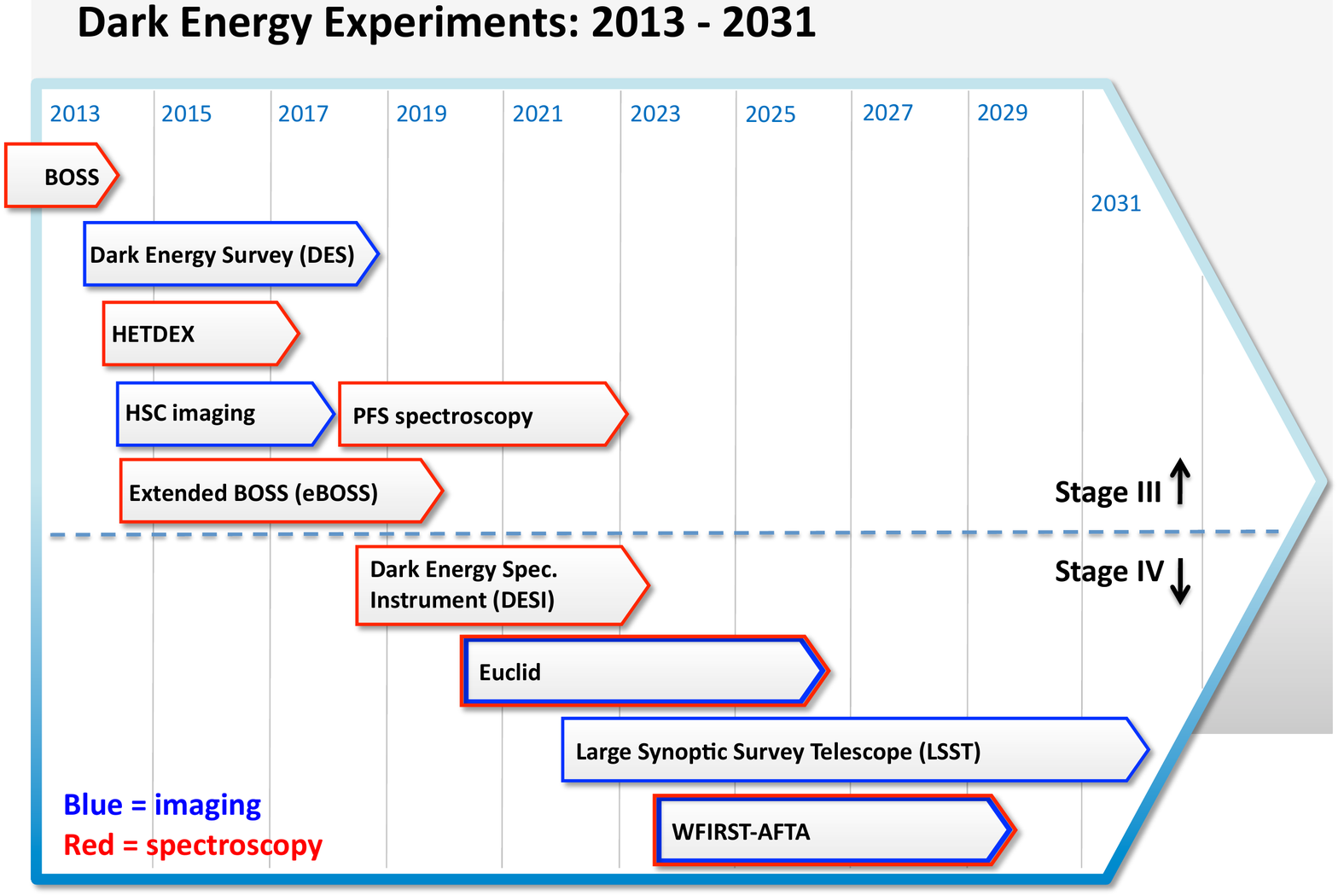}{A timeline of Stage III and Stage IV dark energy
  experiments --- photometric and spectroscopic --- in which U.S.
  scientists are playing an important or leading role.  Most of the
  projects are ground-based with either U.S. leadership (BOSS, DES,
  HETDEX, eBOSS, DESI, LSST) or active participation (HSC, PFS). The
  two space missions are Euclid, led by the ESA with a NASA-sponsored
  team of U.S.~participants, and WFIRST, led by NASA.}

Major gains beyond the current road map of dark energy projects will
require advances on one of a number of fronts.  One potential avenue
is to develop new techniques or probes of cosmology and new physics
that have not yet been developed. Data from many cosmological surveys
have been used for tests not envisioned when those experiments were
first designed, and we anticipate that that trend will
continue~\cite{Jain:2013wgs}.  Another path is to obtain complementary
information that will enhance planned experiments.  For example, a
targeted program of spectroscopic redshift measurements can enhance
the dark energy constraints from LSST compared to its baseline
capabilities~\cite{Zhan:2006gi}.  Finally, new instrumental
capabilities could enable powerful new dark energy experiments to be
conducted at reduced costs. Possible examples include methods that
suppress flux from emission lines from the night sky, which are much
brighter than distant galaxies at infrared wavelengths, or new
detector technologies that would measure photon energy rather than
just quantity.  Modest investments in these avenues may yield large
potential payoffs in the future.

The following sub-sections summarize the physical probes and
measurements that address dark energy science.  Each of these
approaches is detailed further in a separate
document~\cite{Kim:2013ijr,Huterer:2013xky,Jain:2013wgs,Rhodes:2013fyq}.
Taken together, these measurements and the projects that enable them
constitute an all-out attack on the problem of cosmic acceleration.
The program planned over the coming decade guarantees continuing U.S.
leadership in this field.

\subsubsection{Distances}

The relationship between the redshift and distance of an object is one
of the primary tests of the expansion history of the Universe, and
therefore played a key role in the discovery of the accelerating
Universe. The simple graph of the distance scale of the Universe as a
function of redshift, indicating the evidence for cosmic acceleration,
has become an iconic plot in the physical sciences.  The data for this
plot so far come from measurements of Type Ia supernovae and baryon
acoustic oscillations (BAO), and these will be the sources of streams
of data in coming years. DES and LSST will provide an essentially
limitless supply of supernova --- thousands, then hundreds of thousands.
The DES collaboration and LSST-DESC will coordinate the spectroscopic
classification of a fraction of these objects.  The challenge is to
make sufficiently thorough measurements to mitigate systematic
problems, especially those that depend on redshift. Detailed studies
of nearby supernovae are beginning to provide clues for how to do
this. Much would be gained if observations could be made from space,
but a substantial gain will be also achieved if we make ground-based
observations that avoid the atmospheric lines in the near infrared.

The subtle pattern of anisotropy in the CMB, just one part in $10^5$,
is mapped at the two-dimensional boundary of a three-dimensional
feature, the fluctuations in matter density throughout space.  The
counterpart of the oscillations in the CMB power spectrum is a peak in
the correlation between the densities at points separated by 153 Mpc,
measured relative to the scale size of the Universe, which is left
behind by BAO in the early Universe. This very large meter stick can
be observed as far out as redshifts $z = 1.6$ using galaxies as traces
of matter density, and even out to $z = 3$ by observing correlations
in the troughs of spectra from distant
quasars~\cite{Slosar:2013fi}. The current Baryon Oscillation
Spectroscopic Survey (BOSS) is likely to report a distance measurement
soon with 1\% accuracy. The eBOSS survey is designed to extend the
reach of BAO measurements to higher redshifts.  The Stage-IV BAO
experiment, Dark Energy Spectroscopic Instrument (DESI), should
provide more than 30 similarly accurate independent distance
measurements.

\Sfigp{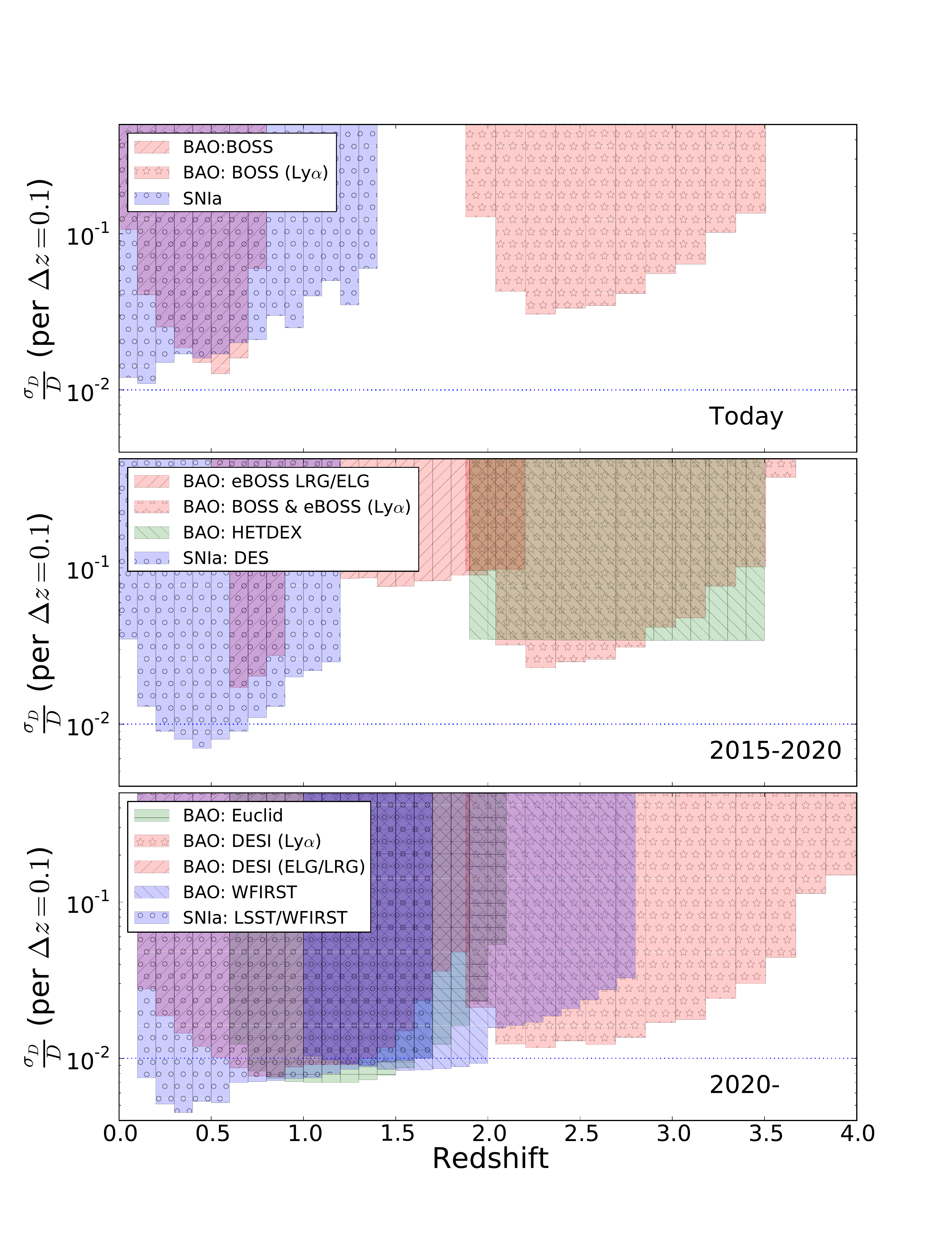}{Current and projected future uncertainties on
  cosmic distance as a function of redshift.}

\Figref{distancekiller} shows the current and projected future
constraints on cosmic distances using these two techniques.  If our
basic understanding is correct, the supernova and BAO measurements
should be in absolute agreement.  The distance--versus--redshift curve
of the Universe is fundamental, and exploring it with completely
different techniques is essential. These stunning measurements will
allow for percent-level determination of the equation of state and
will be extremely sensitive to evolution of the dark energy at earlier
times. By pinning down the distance--redshift relation, they will also
allow for apples-to-apples comparisons of modified gravity versus dark
energy models using the growth of structure.

\subsubsection{Growth of structure}
\label{sec:growth}

The quantity and quality of cosmic structure observations have greatly
accelerated in recent years, and further leaps forward will be
facilitated by imminent projects.  These will enable us to map the
evolution of dark and baryonic matter density fluctuations over cosmic
history.  The way that these fluctuations vary over space and time is
sensitive to several pieces of fundamental physics: the primordial
perturbations generated by GUT-scale physics, neutrino masses and
interactions, the nature of dark matter, and dark energy.  Here we
focus on the last of these, and the ways that combining probes of
growth with those of cosmic distances will pin down the mechanism
driving the acceleration of the Universe.

If the acceleration is driven by dark energy, then distance
measurements provide one set of constraints on $w$, but dark energy
also affects how rapidly structure grows.  Upcoming surveys are
therefore designed to probe $w$ in two distinct ways: direct
observations of the distance scale and the growth of structure, each
complementing the other on both systematic errors and dark energy
constraints.  A consistent set of results will greatly increase the
reliability of the final answer.

\Sfig{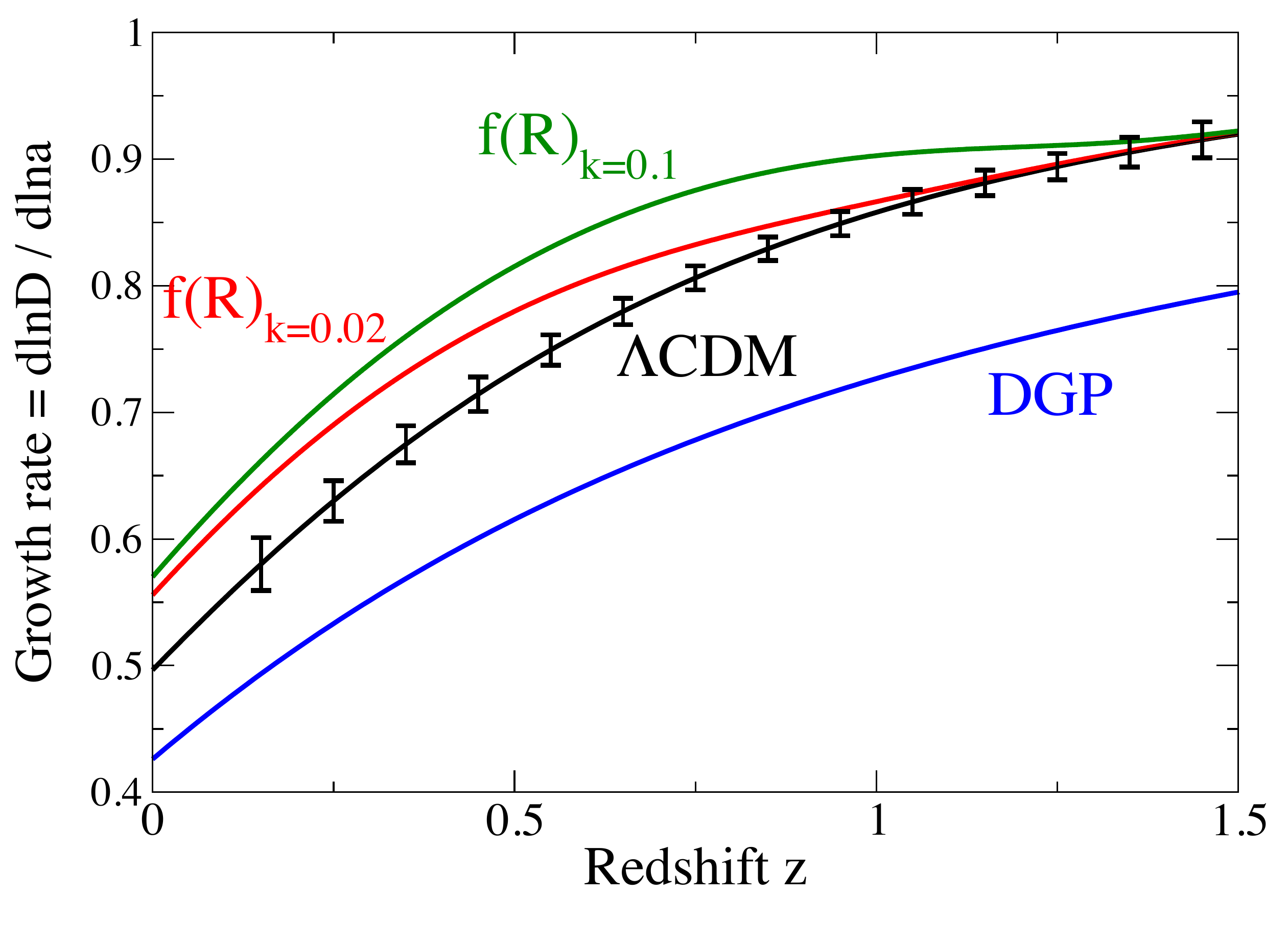}{Constraints on the growth of density fluctuations
  in the Universe with errors projected from DESI. The curves show the
  derivative of the logarithmic growth with respect to logarithmic
  scale factor --- a quantity readily measured from the clustering of
  galaxies in redshift space --- as a function of redshift. We show
  theory predictions for the standard $\Lambda$CDM model, as well as
  for two modified-gravity models, the Dvali-Gabadadze-Porratti (DGP)
  model~\cite{DGP}, and the $f(R)$ modification to Einstein
  action~\cite{0905.2962}. Because growth in the $f(R)$ models is
  generically scale-dependent, we show predictions at wave numbers,
  $k=0.02\,h\,{\rm Mpc^{-1}}$ and $k=0.1\,h\,{\rm Mpc^{-1}}$.  LSST
  projects to impose constraints of similar excellent quality on the
  growth function $D(a)$.  }

If cosmic acceleration is driven by modified gravity, then probes of
structure become even more important.  Generically, modified gravity
models are able to reproduce any expansion history that can be
attained in dark energy models, but at the cost of altering the growth
of structure. How rapidly structure grows is quantified by the
dimensionless {\em growth function} $D(a)$. \Figref{growth_killer}
illustrates how different models make predictions that differ from
those of GR even though the distance--redshift relation
in all these models is identical. The growth of structure then will be
able to distinguish modified gravity from dark energy as an
explanation for the cosmic acceleration.

\Figref{growth_killer} projects constraints from a spectroscopic
survey that measures the local velocities of galaxies by observing
redshift space distortions.  Similarly powerful constraints are
projected from photometric surveys that are dedicated to measuring the
shapes of galaxies and therefore are sensitive to the signal from weak
gravitational lensing.

Achieving these powerful constraints will require both wide field
imaging and spectroscopic redshift surveys, as depicted in
\Figref{Facilities}.  The results will pin down far more than the
equation of state to percent-level accuracy, although this in itself
will be an important clue as to whether the cosmological constant or
an alternative is driving the acceleration of the Universe. We will
also learn whether the equation of state is varying with time and
whether dark energy was relevant at high redshift. The surveys will
probe cosmological perturbations as a function of both length and
time, opening up dozens of possible failure modes for GR-based dark
energy.  If any of the possible deviations from GR predictions is
reliably established, we will have a revolution on our hands.

\subsubsection{Novel probes}
\label{sec:novel}

Surveys enabling the twin probes of distances and structure can
distinguish between modified gravity and dark energy on cosmological
scales. It has become apparent over the last few years that
non-cosmological tests can also play an important role in determining
the mechanism driving the acceleration of the Universe. The basic idea
is that gravity is known to reduce to GR on solar system scales, so
any modified gravity model must have a screening mechanism, wherein
the additional forces operative on large scales are suppressed in the
solar system. Indeed, many of the models have screening built into
them, so the solar system constraints can be naturally satisfied. The
key issue is the nature of the transition from large (modified
gravity) to small (GR) scales, and how that transition can be
observed.

Most screening mechanisms utilize some measure of the mass
distribution of halos, such as the density or the Newtonian potential,
to recover GR well within the Milky Way. This leaves open the
possibility that smaller halos, the outer parts of halos, or some
components of the mass distribution, are unscreened and therefore
experience enhanced forces. For a given mass distribution, unscreened
halos will then have internal velocities and center-of-mass velocities
larger than predicted by GR. Deviations from GR are typically at the
ten percent level, with distinct variations between different
mechanisms in the size of the effect and the way the transition to GR
occurs. It is important to note that observable effects are typically
larger on galaxy scales than on large (cosmological) scales or at high
redshift.  The comparison of dynamical and lensing masses provides a
powerful test that is being implemented on a wide range of scales,
from individual galaxies to large-scale cross-correlations (see
\Secsref{growth}{cross-correlations}).

\Sfig{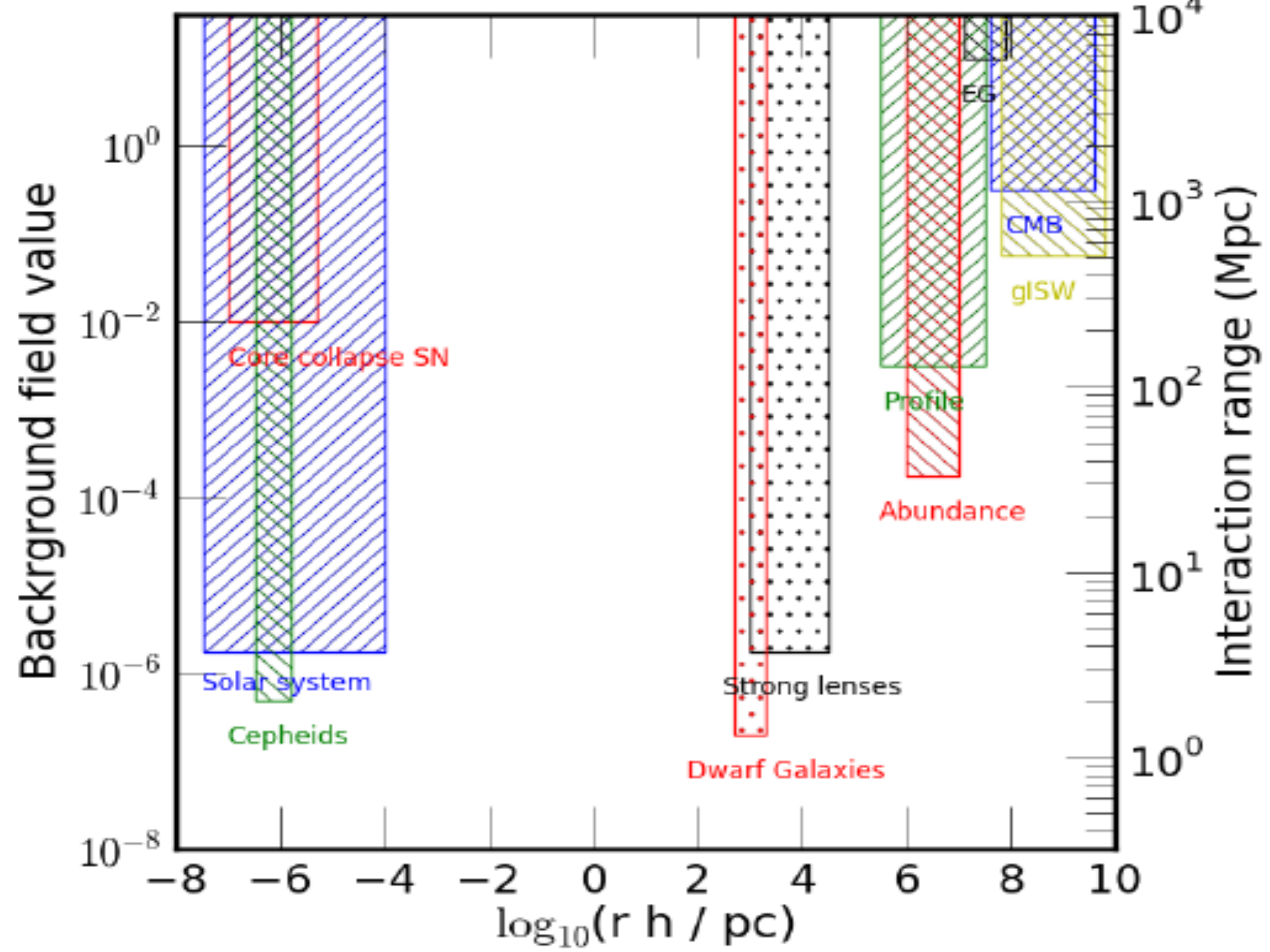}{Astrophysical~\cite{Hu:2007nk,Jain:2012tn,Vikram:2013uba}
  and
  cosmological~\cite{Song:2007da,Giannantonio:2009gi,Schmidt:2009am}
  limits on chameleon theories, in particular $f(R)$ models.  The
  $x$-axis is the distance in units of $h^{-1}$ pc, where $h$
  parametrizes the Hubble constant.  The spatial scale on the $x$-axis
  gives the range of length scales probed by particular experiments.
  The parameter on the $y$-axis is the value of the scalar field
  responsible for acceleration in units of the Planck mass or,
  equivalently, the range of the additional force mediated by this
  field.  The rectangles show excluded regions; the two rectangles
  with dots indicate preliminary results from ongoing work.}

Although the Vainshtein screening mechanism mentioned earlier arises
in some models, the chameleon mechanism, where the extra degree of
freedom is suppressed in high density regions, is prevalent in others.
\Figref{novel} shows a range of tests that have been implemented in
one particular model, $f(R)$ gravity, with chameleon screening. The
screening in this model, as in many others, depends on the value of a
field in the region of interest. The $y$-axis in \Figref{novel}
depicts the value of this field in Planck units divided by a coupling
constant, while the $x$-axis shows that the model has been probed on
scales ranging from the solar system all the way out to cosmological
scales, with all tests to date verifying GR. In the next decade
essentially the entire accessible parameter space of chameleon
theories can be probed using the tests shown in the figure. Tests for
the other important screening mechanism, Vainshtein screening, are at
early stages with potential for rapid progress.

The program of testing gravity theories via these novel probes is in
its infancy, but it is becoming increasingly clear that even modest
investments in non-cosmological observations have enormous potential
to contribute to the cosmic acceleration problem.

\subsubsection{Cross-correlations}
\label{sec:cross-correlations}

To extract the most information possible about dark energy from galaxy
surveys and CMB experiments, scientists will be forced to combine
probes.  The multi-probe approach is not simply a matter of
multiplying uncorrelated likelihoods. For example, gravitational
lensing and large-scale structure are highly correlated probes, with
the lensing signal enhanced behind over-dense regions and the density
enhanced (due to magnification) behind regions with large lensing
signals. The likelihoods are therefore not independent, and there is
significant information contained in the cross-correlations. Beyond
improvement in statistical power, these cross-correlations will enable
us to isolate systematic uncertainties.  The four principal probes ---
supernovae, clusters, lensing, and large-scale structure --- are
correlated with one another in ways that will need to be accounted for
as we strive for percent-level accuracy in dark energy parameters.

The four probes enabled by optical surveys will be supplemented by
data at other wavelengths. One of the most promising for the purposes
of dark energy will come from the oldest probe, the CMB. To date, CMB
experiments have supplemented galaxy surveys mainly by constraining
parameters that would otherwise be degenerate with those that
characterize dark energy. Since the CMB is sensitive to early Universe
physics, it probes dark energy only through the geometric projection
from the surface of last scattering.  The photons from the last
scattering surface, however, experience deflections due to
gravitational lensing and to Compton scattering off free
electrons. These deflections show up as {\em secondary anisotropies},
which often do carry information about dark energy.

Galaxy clusters are perhaps the quintessential example of the value of
observing cosmic phenomena at different wavelengths. One of the key
uses for clusters in dark energy studies is to compare their abundance
above a given mass threshold as a function of redshift with
theoretical predictions. The key uncertainty remains the mass
determination, and it is in this regard that multi-wavelength studies
become particularly important. By observing clusters in optical
surveys, in the microwave frequencies via the Sunyaev-Zel'dovich
effect that follows from Compton scattering off hot electrons in the
clusters, and in the X-ray region when those same electrons emit
radiation, we obtain many possible mass proxies.  Taken together, they
offer a powerful attack on the dominant systematic uncertainties in
dark energy cluster studies. \Figref{sptcl} shows an example of these
different views of a galaxy cluster.

\Sfig{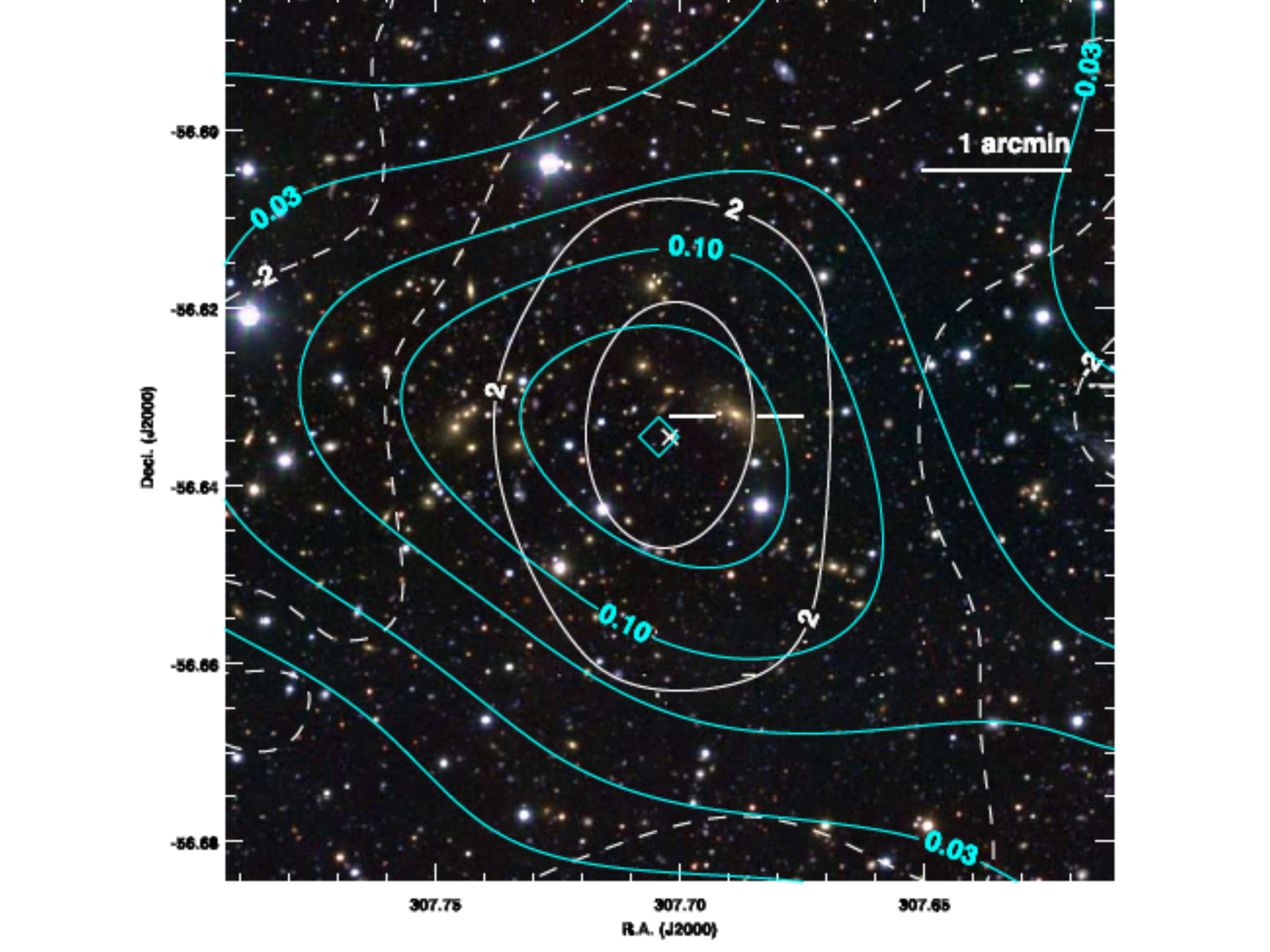}{Map of a galaxy cluster~\cite{High:2012un} using three
  probes: (i) Weak gravitational lensing (blue contours with labels
  showing the projected density $\kappa$); (ii) hot gas as measured by
  the Sunyaev-Zel'dovich distortion of the CMB (white contours with
  labels giving signal to noise); and (iii) galaxies as observed in
  three optical bands (background).}

\subsection{Inflation}
\label{sec:inflation}

Cosmic inflation is the leading theory for the earliest history of the
Universe and for the origin of structure in the Universe.  Current
observations of the large-scale distributions of dark matter and
galaxies in the Universe and measurements of the CMB are in stunning
agreement with the predictions of inflation.  The next generations of
experiments in observational cosmology are poised to explore the
detailed phenomenology of the earliest moments of the Universe.

\Sfig{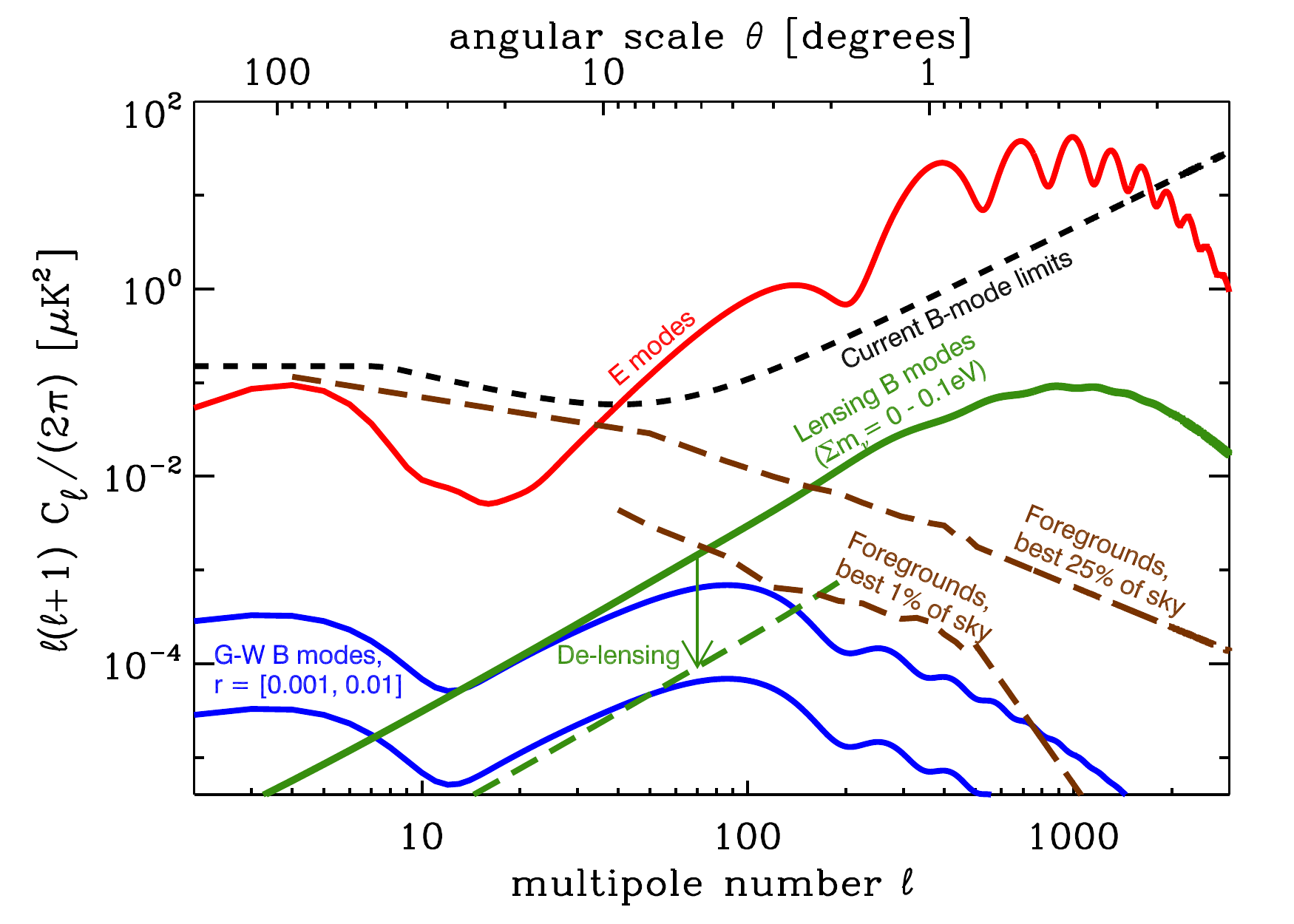}{Expected signal levels for the CMB
  Polarization $E$-mode (solid red), inflationary gravity-wave
  $B$-mode (solid blue), and lensing $B$-mode (solid green) signals.
  The gravitational wave $B$-mode signals are shown for
  tensor-to-scalar ratios of $r=0.001$ (the Stage IV goal) and
  $r=0.01$ (the boundary between small-field and large-field inflation
  models). The lensing $B$-mode signal is shown as a band encompassing
  the predicted signal for values of the sum of neutrino masses $0 \le
  \sum m_\nu \le 0.1 \mathrm{eV}$.  De-lensing by a factor of 4 in
  amplitude is shown schematically by the green arrow, with the
  residual signal at $\ell \le 200$ (where the de-lensing is critical
  to the constraint on $r$) shown by the green, long-dashed line.  The
  black, short-dashed line shows the level of current 95\% upper
  limits on $B$-modes from \textsc{WMAP}, \textsc{BICEP},
  \textsc{QUIET} and \textsc{QUaD} experiments. The brown, long-dashed
  lines show the expected polarized foreground contamination at 95~GHz
  for the cleanest 1\% and 25\% of the sky.}

Although the landscape of possible models for inflation is large, the
theoretical underpinnings are well understood, and we are able to make
concrete predictions for observable quantities.  One key prediction is
the existence of a background of gravitational waves from inflation
that produces a distinct signature in the polarization of the CMB (the
$B$-modes in \Figref{cmb_powspec_for_snowmass}). Under the so-called
Lyth condition~\cite{Lyth:1996im}, all models in which the field
driving inflation varies by an amount of order $m_{\rm Planck}$ will
produce gravitational wave (tensor) fluctuations that are at least $1
\%$ of the amplitude of density-fluctuation (scalar) power in the CMB
($r=0.01$ in the figure).  Definitive evidence as to the presence or
absence of tensor modes with amplitudes at or above this level would
therefore be a window on an infinite sequence of Planck-suppressed
operators.  Such scales require a quantum gravity treatment, and this
will test string-theoretic mechanisms for large field inflation.  If
not detected, these observations at least decide between two broad
classes of models, since they are sensitive to Planck-scale effects.
This motivates the design of a next-generation CMB experiment with the
sensitivity and systematics control to detect such a signal with at
least $5\sigma$ significance, thus ensuring either a detection of
inflationary gravitational waves or the ability to exclude large
classes of inflation models~\cite{Abazajian:2013vfg}.

There are several other ways to probe the physics of inflation.
Inflation generically predicts small deviations from a scale-invariant
spectrum, and current measurements confirm this prediction at the
$5\sigma$ level.  DESI projects to obtain a $15\sigma$ detection,
thereby further reducing the range of allowed models. BOSS, eBOSS, and
DESI will potentially constrain the {\em running} of the primordial
spectrum (deviation from a pure power law) at the 0.2\% level, a
factor of five tighter than current constraints~\cite{Ade:2013uln}.

The non-gaussian distribution (non-gaussianity) of the primordial
perturbations can take many forms. The search for one --- so-called
{\em local} non-gaussianity --- is particularly important because
single-field models of inflation generically predict negligible local
non-gaussianity~\cite{Abazajian:2013vfg}, so any detection will
falsify a large class of models. Planck has placed strong upper limits
on this and other forms of non-gaussianity consistent with these
predictions.  The upcoming surveys eBOSS, DESI, and LSST will
constrain a variety of forms of primordial non-gaussianity on different
spatial scales and be subject to different systematic uncertainties
than the CMB. They will therefore pave the way for even more stringent
bounds on inflationary models.

\subsection{Neutrinos}
\label{sec:neutrinos}

One of the most remarkable aspects of physical cosmology is that the
study of the largest physical structures in the Universe can reveal
the properties of particles with the smallest known cross section, the
neutrinos.  At the simplest level, this cosmological sensitivity to
neutrino properties is due to the fact that the neutrino cosmological
number density is so large as to be second only to CMB photons.  More
specifically, the properties of neutrinos alter the effective energy
density of cosmological radiation and therefore the amplitude, shape
and evolution of matter perturbations, leading to changes in
observables in the CMB anisotropies and in measures of large-scale
structure.

The CMB and large-scale structure measured in galaxy surveys are
sensitive to the sum of the neutrino masses and to $N_{\rm eff}$, the
number of species produced in the early Universe. These observations
are therefore complementary to laboratory probes of neutrinos that
measure mass differences and, potentially, CP violation. CMB
experiments are sensitive enough to the neutrino energy density to
exclude $N_{\rm eff}=0$ at more than $10\sigma$; \ie, these
experiments have already indirectly detected the cosmic
neutrinos. Together with existing data on the redshift-distance
relation from galaxy surveys like BOSS, these experiments also place a
stringent upper limit on the sum of the neutrino masses, currently
around 0.23 eV~\cite{Ade:2013zuv}.

A global fit to solar and atmospheric neutrino flavor oscillations in
the standard 3-generation model determines two mass differences. A
third parameter, which can be taken as either the sum of the masses or
the lightest mass, is, therefore, unknown. Due to a sign ambiguity in
one of the mass differences, there are two discrete possibilities,
normal and inverted hierarchy, for the relationship between these two
parameters.  As indicated in \Figref{numass_combine2}, upcoming CMB
experiments and large-scale structure surveys will unambiguously
detect the sum of the masses if the hierarchy is inverted and will
likely do so at greater than $3\sigma$ significance, even if nature
has chosen the normal hierarchy.  Measures of the power spectrum from
DESI, combined with Planck data, could improve the current constraint
on $\sum m_\nu$ to 17 meV, and a Stage IV CMB survey combined with BAO
measurements from DESI could achieve similar
sensitivity~\cite{Abazajian:2013oma}.  \Figref{numass_combine2}
highlights another aspect of the complementarity of laboratory
experiments and cosmological probes: If the sum of the masses is found
to be $0.15$ eV by the cosmological probes, then it will take the lab
experiments to determine the lightest neutrino mass by identifying
whether the hierarchy is normal or inverted.

\Sfig{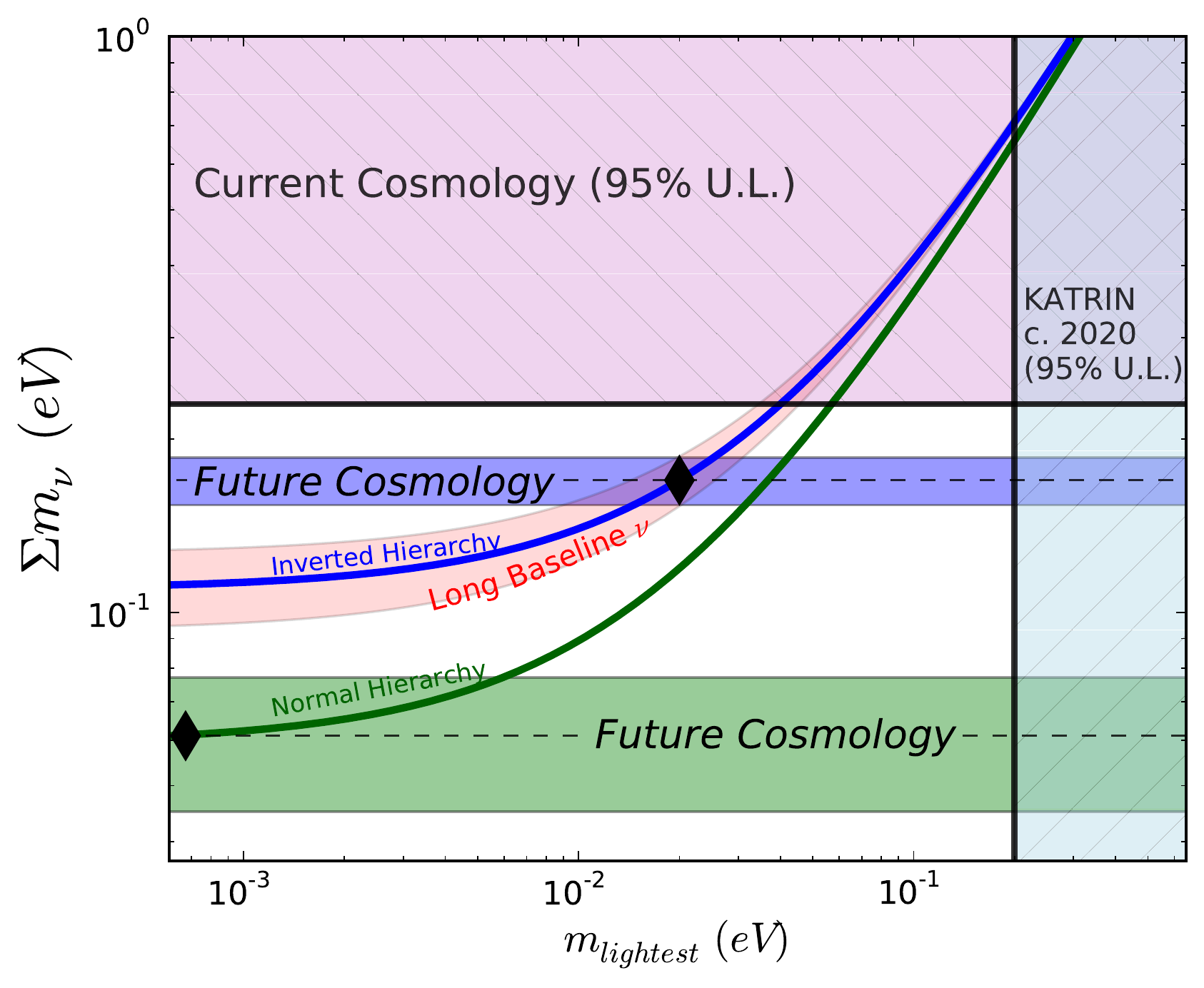}{Current constraints and forecast sensitivity of
  cosmology to the sum of neutrino masses. In the case of an
  ``inverted hierarchy,'' with an example case marked as a diamond on
  the upper curve, future combined cosmological constraints would have
  a very high-significance detection, with $1\sigma$ error shown as a
  blue band.  In the case of a normal neutrino mass hierarchy with an
  example case marked as a diamond on the lower curve, future cosmology
  would still detect the lowest $\sum m_\nu$ at greater than
  $3\sigma$.}

Short-baseline neutrino oscillation results hint at a richer neutrino
sector than three active neutrinos participating in flavor
oscillations, with one or more sterile flavors also
participating~\cite{Aguilar:2001ty,AguilarArevalo:2010wv,Kopp:2011qd,Giunti:2011gz}.
Future CMB experiments will achieve a $1\sigma$ error of $\Delta
N_{\rm eff} = 0.027$, which will complement future sterile neutrino
searches and inform model building, since sterile neutrinos might be
detected in the laboratory even though they were not produced in the
early Universe and vice versa.

\subsection{Dark energy and CMB conclusions}

Cosmological surveys are sensitive to fundamental physics. To date,
this has led to the discovery of the accelerating Universe, strong
evidence for an epoch of early acceleration near the GUT scale, the
indirect detection of the cosmic neutrino background, and the most
compelling evidence for non-baryonic dark matter. However, surveys to
date have measured only a fraction of all information available. If
the Universe were contained in an area the size of the United States,
galaxy maps so far would have surveyed the city of Birmingham,
Alabama. CMB experiments have provided low-noise maps of the
temperature down to angular scales of order a tenth of a
degree. Strategic, valuable information remains unmined in
higher-resolution temperature maps and the virtually uncharted
polarization field. We have outlined the projections for how this
extra information will constrain dark energy, neutrinos, and
inflation, and these projections are extraordinary. But even they
ignore the very real possibility that future experiments on the Cosmic
Frontier will do just what their predecessors have done: Discover
something fundamentally new!

The community has rallied behind previous
reports~\cite{Albrecht:2006um,Albrecht:2009ct,rocky3} which are
consistent with the current consensus to support the following key
steps:

\begin{itemize}

\item {\bf Remain a leader in dark energy research.}  The U.S. played
  the leading role in discovering the acceleration of the Universe, as
  was recognized by the 2011 Nobel Prize. The acceleration remains a
  mystery, whose solution may usher in a revolution in either our
  theory of gravity or our understanding of particle
  physics. Different classes of theories make different predictions
  for the growth of structure, given a redshift-distance relation. A
  combination of spectroscopic and photometric surveys can determine
  both distances and structure growth, and so will help pinpoint the
  new physics driving the acceleration of the Universe. The current
  suite of surveys, Stage III, will be the first to implement the
  vision of multiple probes and small systematic uncertainties.  This
  vision will be realized fully with the Stage IV surveys (LSST and
  DESI) when they reach the level where exquisite-precision dark
  energy constraints from different probes, in some cases approaching
  the cosmic variance limit, can be checked for consistency.
  Therefore, the community strongly supports continuing the program of
  Stage III and Stage IV dark energy experiments, and moving forward
  as quickly as possible with the construction of LSST and DESI.

\item {\bf Build a generation IV CMB polarization experiment.}  CMB
  experiments can measure the sum of the neutrino masses and the
  energy scale of inflation, as well as constrain exotic physics such
  as early dark energy and extra neutrino species. After the current
  generation of small-scale experiments complete data-taking near the
  end of the decade, the community understands that the next
  generation experiment --- one that can pin down neutrino masses and
  the scale of large field inflationary models --- requires a
  nationwide coherent effort. Moving from thousands to hundreds of
  thousands of detector elements will require the involvement of the
  national laboratories working together with the university
  community.

\item {\bf Extend the reach.}  With small additional investments, the
  dark energy program can be augmented in three important ways. A
  targeted spectroscopic campaign designed to optimize and calibrate
  methods of redshift estimation from imaging surveys can enhance
  their science returns beyond their nominal
  capabilities~\cite{Newman:2013cac}. Second, a continued investment
  in instrumentation R\&D will allow the community to do more science
  for less money and to be ready to capitalize on future discoveries.
  Finally, a suite of novel probes of gravity and dark energy can
  discover signatures of modified gravity and new physics in the dark
  sector. We have the opportunity to catalyze the next generation of
  tests by supporting work at the interface of theory, simulation and
  data analysis, and making small enhancements to the dark energy
  survey program.

\end{itemize}

\section{Cosmic particles and fundamental physics} 

Over the past decade we have witnessed a revolution in our
understanding of the high-energy Universe.  This includes several key
discoveries.
\begin{itemize}
\item Supernova remnants have been shown to be a source of galactic cosmic
  rays~\cite{Ackermann:2013wqa}.
\item Very high energy neutrinos that are likely to be astrophysical
  in origin have been observed~\cite{Aartsen:2013bka}.
\item The GZK suppression in the cosmic-ray flux above $3 \times
  10^{19}$ eV has been observed~\cite{Abbasi:2007sv,Abraham:2008ru}.
\item The positron fraction of the cosmic rays has been measured up to
  300 GeV, providing solid evidence for a high-energy primary source
  of positrons in the galaxy, either from dark matter annihilation or
  astrophysical processes~\cite{Adriani:2008zr,Aguilar:2013qda}.
\item Many sites of astrophysical particle acceleration have been
  directly observed, from supermassive black holes and merging neutron
  stars, to rapidly spinning neutron stars and supernova remnants in
  our galaxy~\cite{Aharonian:2008zza}.
\end{itemize}
These discoveries have been driven by the current generation of
experiments: the IceCube neutrino detector at the South Pole, the
Fermi gamma-ray observatory and the PAMELA and AMS experiments
orbiting the Earth, the High Resolution Fly's Eye and Pierre Auger
Observatory ultra-high-energy cosmic ray experiments, and the HESS,
VERITAS, MAGIC and Milagro experiments in very-high-energy gamma rays.

Looking forward, a new generation of instruments with greater
sensitivity and higher resolution holds the promise of making large
advances in our understanding of astrophysical processes and the
fundamental physics studied with astrophysical accelerators.  The
goals for the coming decade are:
\begin{itemize}
\item Determine the origin of the highest energy particles in the
  Universe and understand the acceleration processes at work
  throughout the Universe.
\item Measure particle cross sections at energies unattainable in
  Earth-bound accelerators.
\item Measure the highest energy neutrinos that arise from
  interactions of the ultra-high-energy cosmic rays with the CMB.
\item Measure the extragalactic background light at wavelengths
  between 1 and 100 $\mu$m to understand the star formation history of
  the Universe.
\item Measure the mass hierarchy of the neutrinos.
\item Search for physics beyond the Standard Model encoded in cosmic
  messengers as they cross the Universe.
\item Understand the origins of the matter-antimatter asymmetry of the
  Universe.
\item Probe the fundamental nature of spacetime.
\end{itemize}

In many of these areas, future progress will depend upon either the
detailed understanding of particle acceleration in the Universe or the
development of methods for controlling systematic errors introduced by
our lack of understanding of these processes.  High-resolution gamma
ray measurements (spectral, angular, and temporal) of many objects and
classes of objects are needed to find the source-invariant physics
that is the signal for physics beyond the Standard Model.  Such
measurements, in conjunction with measurements at other wavelengths
and with measurements of cosmic rays, neutrinos, and gravity waves,
will enable us to understand Nature's particle accelerators. We
propose several recommendations to achieve these goals.
\begin{itemize}
\item {\bf Significant U.S. participation in the CTA
  project~\cite{Consortium:2010bc}.}  U.S. scientists developed the
  imaging atmospheric Cherenkov technique.  Continued leadership in
  this area is possible with the development of novel telescope
  designs.  A U.S. proposal to more than double the number of
  mid-scale telescopes would result in a sensitivity gain of 2 to 3,
  significantly improving the prospects for the indirect detection of
  dark matter, understanding particle acceleration processes, and
  searching for other signatures of physics beyond the Standard Model.
\item {\bf Simultaneous operation of Fermi, HAWC, and VERITAS.}
  Understanding particle acceleration and separating astrophysical
  processes from physics beyond the Standard Model requires
  observations over a broad energy range.  The above three instruments
  will provide simultaneous coverage from 30 MeV to 100 TeV.  HAWC and
  VERITAS will simultaneously view the same sky, enabling prompt
  follow-up observations of transient phenomena.
\item {\bf Construction of the PINGU neutrino
  detector~\cite{IceCube:2013aaa} at the South Pole.}  U.S. scientists
  have been leaders in the field of high-energy neutrino observations.
  PINGU, by densely instrumenting a portion of the IceCube Deep Core
  array, will lower the energy threshold for neutrinos to a few GeV.
  This will allow for a measurement of the neutrino mass hierarchy
  using atmospheric neutrinos.
\item {\bf Continued operation of the Auger and Telescope Array (TA)
  air shower arrays with upgrades} to enhance the determination of the
  composition and interactions of cosmic rays near the energy of the
  GZK suppression, {\bf and flight of the JEM-EUSO
    mission~\cite{TheJEM-EUSO:2013vea}} to extend observations of the
  cosmic ray spectrum and anisotropy well beyond the GZK region.
\item {\bf Construction of a next-generation ultra-high-energy GZK
  neutrino detector} either to detect GZK
  neutrinos~\cite{Berezinsky:2002nc} and constrain the
  neutrino-nucleon cross section at these
  energies~\cite{Klein:2013xoa}, or exclude all but the most
  unfavorable parts of the allowed parameter space.
\end{itemize}

\subsection{Ultra-high-energy cosmic rays}

HiRes~\cite{Abbasi:2007sv}, Auger~\cite{Abraham:2010mj}, and
TA~\cite{AbuZayyad:2012ru} have established the existence of a
suppression of the spectrum at the highest energies (above
$\sim 3 \times 10^{19}$ eV), as predicted by Greisen, Zatsepin, and
Kuzmin (GZK) in 1966~\cite{Greisen:1966jv,Zatsepin:1966jv}.  The GZK
suppression is an example of the profound links between different
regimes of physics, connecting the behavior of the highest-energy
particles in the Universe to the CMB, and can be explained by the
sub-GeV-scale physics of photo-pion production occurring in the
extremely boosted relativistic frame of the cosmic ray.  A similar
phenomenon occurs for primary nuclei due to excitation of the giant
dipole resonance, resulting in photo-disintegration.  For iron nuclei,
this occurs at about the same energy per particle as the photo-pion
process does for protons.

The composition of cosmic rays and their interactions with air nuclei
may be probed by studies of the depth of shower maximum, $X_{\rm
  max}$~\cite{Ellsworth:1982kv}.  The mean value of $X_{\rm max}$
rises linearly as a function of the logarithm of the cosmic ray's
energy, and depends on the nature of the primary particle, the depth
of its first interaction, and the multiplicity and inelasticity of the
interactions as the shower evolves.  Lower energy observations of
$X_{\rm max}$ indicate that the composition becomes lighter as the energy
increases toward $\sim10^{18.3}$ eV~\cite{Bird:1993yi}, which suggests
that extragalactic cosmic rays are mainly protons.  However, at higher
energies the Auger Observatory's high-quality, high-statistics sample
exhibits the opposite trend, along with a decreasing spread in $X_{\rm
  max}$ with increasing energy~\cite{Abraham:2010mj,Abreu:2013env}.
Using current simulations and hadronic models tuned with LHC forward
data, this implies the composition is becoming gradually heavier above
$10^{18.5}$ eV.  A trend toward heavier composition could reflect the
apparent GZK suppression being in fact the endpoint of cosmic
acceleration, in which there is a maximum magnetic rigidity for
acceleration, resulting in heavy nuclei having the highest energy per
particle.

Cosmic rays can be used to probe particle physics at energies far
exceeding those available at the LHC.  An alternative explanation for
the observed behavior of $X_{\rm max}$ is a change in particle
interactions not captured in event generators tuned to LHC data.
Auger measurements using three independent methods find that these
models do not describe observed showers well.  For example, the
observed muon content of showers measured in hybrid events at Auger is
a factor 1.3 to 1.6 larger than predicted~\cite{AugerMuonICRC13}.  TA
also observes a calorimetric energy that is about 1.3 times higher
than that inferred from their surface detector using these models.  An
example of a novel phenomenon that may explain these observations is
the restoration of chiral symmetry in QCD~\cite{Farrar:2013sfa}.

A critical step in fully understanding the $X_{\rm max}$ observations
is to identify and correct the deficiencies in the beyond-LHC physics
used in modeling showers.  This requires continued operation of
current hybrid detectors such as Auger and TA, with upgrades to enable
improved multi-parameter studies of composition and interactions on a
shower-by-shower basis.  Enhancements of the surface detectors are
particularly valuable because of the tenfold higher duty cycle than
for fluorescence or hybrid operation.

Observations from space can extend studies of the spectrum and
anisotropy beyond the GZK region with high statistics.  Measurements
at ground-based observatories hint that cosmic ray arrival directions
may be correlated with the local distribution of
matter~\cite{Kampert:2012vh,TinyakovTAanisoICRC13}, but higher
statistics trans-GZK observations are required to identify sources. In
addition, answering the question of whether the spectrum flattens
again above the GZK suppression or continues to fall will distinguish
between the GZK and acceleration limit scenarios.  The JEM-EUSO
mission has an instantaneous collecting area of $\sim$40 times that of
existing ground-based detectors~\cite{TheJEM-EUSO:2013vea} and, taking
duty cycle into account, will increase the collecting area above the
GZK suppression energy by nearly an order of magnitude.

\subsection{Neutrinos}

IceCube has recently reported the detection of two neutrinos with
energies above 1 PeV and 26 events above 30 TeV, with characteristics
that point to an astrophysical origin~\cite{Aartsen:2013bka}.  These
exciting results herald the beginning of the era of high-energy
neutrino astronomy and initiate the study of ultra-long baseline
high-energy neutrino oscillations.  Neutrino data complement
observations of cosmic rays and gamma rays due to the origin of
neutrinos in the decays following high-energy hadronic interactions
and their weak couplings.  Several acute issues in particle physics
and astrophysics can be addressed by neutrino experiments.

{\bf GZK neutrinos}. GZK neutrinos are created in the weak decays of
the mesons and neutrons produced in the interaction of
ultra-high-energy cosmic rays with the CMB~\cite{Berezinsky:2002nc}.
The production of these neutrinos takes place via well-known physics
at high Lorentz boost, so robust predictions of the neutrino flux can
be made.  The flux depends on the composition of the primary cosmic
rays (being lower for a heavier composition) and the evolution of the
cosmic ray source density with redshift.  Unlike many searches, there
is a lower limit on the expected flux.  Current detectors such as
IceCube, Auger, RICE, and ANITA have begun to probe the highest
predicted values of the neutrino flux.  Next-generation experiments
such as ARA, ARIANNA, and EVA can increase our sensitivity by about
two orders of magnitude, and will either detect GZK neutrinos or
exclude much of the parameter space.  If GZK neutrinos are detected,
the event rate as a function of zenith angle can be used to measure
the neutrino-nucleon cross section and constrain models with enhanced
neutrino interactions at high energy~\cite{Anchordoqui:2005pn}.

{\bf Atmospheric neutrinos and the neutrino mass hierarchy}.  PINGU is
a proposed high-density infill of the IceCube detector with a reduced
energy threshold of a few GeV, employing the rest of IceCube as an
active veto.  PINGU has sensitivity to atmospheric $\nu_{\mu}$ over a
range of values of the length-to-energy ratio $L/E$ spanned by the
variation in the distance to the production region as a function of
zenith angle and the energy spectrum of atmospheric neutrinos.
Preliminary studies of PINGU indicate that over these values of $L/E$,
atmospheric $\nu_{\mu}$ oscillations can be used to determine the
neutrino mass hierarchy with 3--5$\sigma$ significance, given two
years of data with a 40-string detector~\cite{IceCube:2013aaa}.

{\bf Supernova neutrinos}.  The measurement of the time evolution of
the neutrino energy and flavor spectrum from supernova bursts has the
potential to revolutionize our understanding of neutrino properties,
supernova physics, and measure or tightly constrain non-standard
interactions.  Collective oscillations of neutrinos leaving the
neutron star surface imprint distinct signatures on the time evolution
of the neutrino spectrum that depend on the neutrino mass hierarchy
and the mixing angle $\theta_{13}$.  Although the spectral
distributions of anti-electron neutrinos for the normal and inverted
mass hierarchies are not so distinct, the electron neutrino spectra
are dramatically different~\cite{Duan:2006an}.  Therefore, a large
underground liquid argon detector, such as the Long Baseline Neutrino
Experiment (LBNE)~\cite{Adams:2013qkq}, which is predominantly
sensitive to electron neutrinos, could determine the neutrino mass
hierarchy if a supernova occurred within our Galaxy.

\subsection{Gamma rays}
\label{sec:gammarays}

High-energy gamma rays provide a unique view into the most extreme
environments in the Universe, probing particle acceleration processes
and the origin of the galactic and extragalactic cosmic rays.  Active
galactic nuclei (AGN), supermassive black holes emitting jets of
highly relativistic particles along their rotation axis, have been
shown to be sites of particle acceleration~\cite{Aharonian:2008zza}.
Outstanding issues in the acceleration processes include: the nature
of the accelerated particles (hadronic or leptonic), the role of shock
acceleration versus magnetic reconnection, and the formation and
collimation of astrophysical jets~\cite{Boettcher:2006pd}.  Answers to
these questions will come from higher resolution measurements in the
GeV--TeV regime, multi-wavelength campaigns with radio, X-ray, and
gamma-ray instruments, and multi-messenger observations with gamma
rays, ultra-high energy cosmic rays, neutrinos, and potentially
gravitational waves. Understanding these extreme environments and how
they accelerate particles is of fundamental interest.  In addition,
these high-energy particle beams, visible from cosmologically
interesting distances, allow us to probe fundamental physics at scales
and in ways that are not possible in Earth-bound particle
accelerators.

Recently, Fermi~\cite{Ackermann:2013wqa} and
AGILE~\cite{Giuliani:2011nx} measured the energy spectra of the
supernova remnants W44 and IC44.  The decrease in the gamma-ray flux
below the pion mass in these sources is clear evidence for hadronic
acceleration.  This is the clearest evidence to date that some
galactic cosmic rays are accelerated in supernova remnants.  The
detection of high-energy neutrinos from a cosmic accelerator would be
a smoking-gun signature of hadronic acceleration.  In the absence of
multi-messenger signals, multi-wavelength energy spectra (X-ray to
TeV) can test both leptonic models, where the high-energy emission is
derived from inverse Compton scattering of the X-ray synchrotron
emission, and hadronic models, where gamma rays result from pion
decays or proton synchrotron radiation.

{\bf Backgrounds to dark matter searches}. Understanding the origin of
the galactic very high energy gamma rays is critical in the
interpretation of some signatures of dark matter annihilation.  The
recent results from PAMELA~\cite{Adriani:2008zr} and
AMS~\cite{Aguilar:2013qda} of the increasing positron fraction with
energy is a clear signal that the current model of secondary
production and transport through the galaxy is not correct.  There are
three potential explanations for this signal: a new astrophysical
source of positrons~\cite{Hooper:2008kg}, modified propagation of
cosmic rays or secondary production in the
source~\cite{Zirakashvili:2011sc}, or dark matter
annihilation~\cite{Turner:1989kg}.  An astrophysical source of
positrons, pulsar wind nebula, is now known to also lead to an
increasing positron fraction at high energies.  Observations of the
Geminga pulsar wind nebula in the TeV band by
Milagro~\cite{Abdo:2009ku} have been used to normalize the flux of
positrons in our local neighborhood from this source.  The calculated
positron fraction is an excellent match to the
data~\cite{Yuksel:2008rf}.  Similarly, in searching for dark matter
signatures from the GC or galaxy clusters, we must understand and
measure the more standard astrophysical processes that may lead to
signatures that are similar in nature to those expected from dark
matter annihilation.

{\bf Extragalactic background light}. The extragalactic background
light (EBL) pervades the Universe.  It is the sum of all the light
generated by stars and the re-radiation of this light in the infrared
band by dust~\cite{2011MNRAS.410.2556D}, and it is therefore sensitive
to the star formation history of the Universe.  In addition to
advancing our understanding of particle acceleration and astrophysical
backgrounds for dark matter searches, the intense gamma-ray beams
generated by AGN and gamma-ray bursts can be used to probe the
intervening space and search for physics beyond the Standard
Model. Future studies may measure the EBL, use the EBL to measure the
intergalactic magnetic fields, and search for axion-like particles
(ALPs).  The production of electron-positron pairs from photon-photon
scattering of the EBL with high-energy gamma rays leads to an
energy-dependent attenuation length for high-energy gamma
rays~\cite{Gould:1967zza}.  At the same time, cosmic rays, if they are
accelerated by AGN, will interact with the EBL and the CMB along the line
of sight and generate secondary gamma rays at relatively close
distances to the observer~\cite{Aharonian:1993vz}.  This absorption,
with the inclusion of secondary gamma rays, can be used to measure or
constrain the EBL~\cite{2011MNRAS.410.2556D}.  Lower limits on the EBL
can be established from galaxy counts.  An inconsistency between these
lower limits and the measurements of the TeV spectra from AGN would be
a signature of new physics.  Such a discrepancy could be explained
either by the secondary production of gamma rays from cosmic rays
produced at AGN or by photon-ALP mixing mediated by the intergalactic
magnetic fields.

{\bf Intergalactic magnetic fields}. The origin of the galactic
magnetic fields remains a mystery.  Although astrophysical dynamos can
efficiently amplify pre-existing magnetic fields, the generation of a
magnetic field is difficult~\cite{Kandus:2010nw}. The strength of the
magnetic fields in the voids between galaxy clusters should be similar
to the primordial magnetic field. Measurements of AGN spectra and time
delays in the GeV--TeV region have been used to set both lower and
upper bounds on the strength of the intergalactic magnetic
field~\cite{Plaga:1994hq}.  Current bounds are model-dependent and
span a large range of values for the magnetic field.  Improved
measurements of the EBL, the measurement of variability from distant
AGN, improved determinations of the energy spectra, and understanding
of the intrinsic AGN spectra are needed to significantly improve these
limits.

{\bf Tests of Lorentz invariance violation}.  Experimentally probing
Planck-scale physics, where quantum gravitational effects become
large, is challenging.  A unique signature of such effects would be
the violation of Lorentz invariance --- a natural, though not necessary,
property of theories of quantum gravity~\cite{Solodukhin:2011gn}.
Short, intense pulses of gamma rays from distant objects, such as
gamma-ray bursts and active galaxies, provide a laboratory in which to
search for small, energy-dependent differences in the speed of light.
Current limits have reached the Planck scale if the energy dependence
of the violation is linear~\cite{2009Natur.462..331A}, and
$6.4\times10^{10}$ GeV if the violation is
quadratic~\cite{HESS:2011aa} with energy.  Future instruments could
improve these limits by at least factors of 10 and 50, respectively,
and significantly increase the sample size used to search for these
effects.

{\bf TeV gamma-ray instruments}. Ground-based TeV instruments fall
into two classes: extensive air shower (EAS) arrays, capable of
simultaneously viewing the entire overhead sky, and imaging
atmospheric Cherenkov telescopes (IACTs), pointing instruments with
high sensitivity and resolution.  Current IACTs are VERITAS, MAGIC,
and HESS, while the HAWC (under construction) and Tibet arrays are the
only operating EAS arrays.  CTA, the next-generation experiment, will
consist of a large array of IACTs with roughly an order of magnitude
greater sensitivity than current instruments.  It is expected to
discover over 1000 sources in the TeV band~\cite{Acharya:2013sxa}.
The U.S. portion of the collaboration is proposing to more than double
the number of mid-sized telescopes to over 50 using a novel optical
design that will improve the sensitivity of CTA by a factor of 2-3 (a
result of improved angular resolution of the new telescopes and an
increase in the number of telescopes).

\subsection{Baryogenesis}

According to standard cosmology, the current preponderance of matter
arises during the very early Universe from an asymmetry of about one
part per hundred million between the densities of quarks and
antiquarks. This asymmetry must have been created after inflation by a
physical process known as baryogenesis.  Baryogenesis requires
extending the Standard Model of particle physics via some new physics
that couples to Standard Model particles and is important during or
after the end of inflation. Constraints on the inflation scale and on
the reheat temperature at the end of inflation give an upper bound on
the relevant energy scale for baryogenesis. The new physics must
violate CP, as the CP violation in the Standard Model is
insufficient~\cite{Gavela:1994dt}. There are a very large number of
theoretical baryogenesis proposals.

{\bf Leptogenesis}. Theoretically, baryon number violation at high
temperatures is rapid in the Standard Model via nonperturbative
electroweak processes known as sphalerons. Sphalerons conserve the
difference between baryon and lepton numbers, leading to the idea of
leptogenesis~\cite{Fukugita:1986hr}. The decay of very heavy neutrinos
in the early Universe could occur in a CP-violating way, creating a
lepton asymmetry that the sphalerons convert into a baryon
asymmetry. Whether and how leptogenesis could occur can be clarified
by Intensity Frontier experiments that will probe CP violation and
lepton number violation in neutrino physics~\cite{DiBari:2012fz};
Cosmic Frontier experiments, which may, for example, probe the high
inflation scale required in most leptogenesis proposals; and Energy
Frontier experiments~\cite{Pilaftsis:2003gt}, which may discover new
particles, with important implications for leptogenesis.

{\bf Affleck-Dine baryogenesis}. In supersymmetric theories,
condensates of the scalar partners of quarks and leptons have
relatively low energy density and are likely to be present at the end
of inflation~\cite{Affleck:1984fy}. The subsequent evolution and decay
of the condensates can produce the baryon asymmetry, and in some
models, the dark matter~\cite{Kusenko:1997si,Kusenko:1997vp}. The dark
matter could be macroscopic lumps of scalar quarks or leptons known as
Q-balls, which have unusual detection signatures in Cosmic Frontier
experiments~\cite{Kusenko:2005du}. A critical test of this theory is
the search for supersymmetry, primarily via collider experiments.
Instruments such as HAWC and IceCube will be sensitive to extremely
low fluxes of Q-balls --- over two orders of magnitude lower than
current limits.  Although the non-observation of Q-balls cannot
exclude the Affleck-Dine mechanism, the observation of a Q-ball would
have a profound impact on our understanding of the Universe.

{\bf Electroweak baryogenesis}. New particles beyond the Standard
Model that are coupled to the Higgs boson can provide a first-order
phase transition for electroweak symmetry breaking, which proceeds via
nucleation and expansion of bubbles of broken phase. CP-violating
interactions of particles with the expanding bubble walls can lead to
a CP-violating particle density in the symmetric phase, which can be
converted by sphalerons into the baryon
asymmetry~\cite{Cohen:1992zx}. The baryons then enter the bubbles of
broken phase where the sphaleron rate is negligible. This scenario can
be definitively tested by searches for new particles in high-energy
colliders and nonvanishing electric dipole moments for the neutron and
for atoms. Non-standard CP violation in $D$- and $B$-meson physics may
appear in some models. A non-standard Higgs boson self-coupling is a
generic consequence. The first-order phase transition in the early
Universe can show up via relic gravitational waves.

Other experiments that can shed light on baryogenesis include searches
for baryon number violation. Neutron--antineutron oscillations violate
the difference between baryon and lepton numbers, and most
leptogenesis scenarios will not work if such processes are too
rapid. Proton decay would provide evidence for grand unification,
which would imply the existence of heavy particles whose decay could
be responsible for baryogenesis, and which is a feature of some
leptogenesis and supersymmetric models.

\subsection{Fundamental nature of spacetime}

Quantum effects of spacetime are predicted to originate at the Planck
scale. In standard quantum field theory, their effects are strongly
suppressed at experimentally accessible energies, so spacetime is
predicted to behave almost classically, for practical purposes, in
particle experiments. However, new quantum effects of geometry
originating at the Planck scale from geometrical degrees of freedom
not included in standard field theory may have effects on macroscopic
scales~\cite{Verlinde:2010hp} that could be measured by laser
interferometers, such as the holometer~\cite{Hogan:2012ib}. In
addition, cosmic particles of high energies can probe departures from
Lorentz invariance, as discussed in \Secref{gammarays}, and the
existence of extra dimensions on scales above the LHC.

\bibliography{bibCFfinal.bib}


\end{document}